\documentclass[aps,prd,superscriptaddress,9pt,onecolumn,showpacs,showkeys]{revtex4}
\usepackage{amssymb}
\usepackage{amsmath}
\usepackage{amsfonts}
\usepackage{graphicx, float}
\usepackage{graphicx, epsfig}
\usepackage{color}
\usepackage{enumerate}

\newcommand{\beq}{\begin{equation}}
\newcommand{\eeq}{\end{equation}}
\newcommand{\bea}{\begin{eqnarray}}
\newcommand{\eea}{\end{eqnarray}}

\begin{document}

\title{Black Remnants from  T-Duality}

\author{Behnam Pourhassan}\email{b.pourhassan@umz.ac.ir}
\affiliation{Iran Science Elites Federation, Tehran, Iran.}
\affiliation{Canadian Quantum Research Center 204-3002 32 Ave Vernon, BC V1T 2L7 Canada.}
\author{Salman Sajad Wani}\email{salmansajadwani@gmail.com}
\affiliation{Canadian Quantum Research Center 204-3002 32 Ave Vernon, BC V1T 2L7 Canada.}
\affiliation{Department of Physics, University of Kashmir,  Kashmir, India.}
\author{Mir Faizal}\email{mirfaizalmir@googlemail.com}
\affiliation{Department of Physics and Astronomy, University of Lethbridge, Lethbridge, Alberta, T1K 3M4, Canada.}
\affiliation{Irving K. Barber School of Arts and Sciences, University of British Columbia, Kelowna, British Columbia, V1V 1V7, Canada.}
\affiliation{Canadian Quantum Research Center 204-3002 32 Ave Vernon, BC V1T 2L7 Canada.}

\begin{abstract}
In this paper, we will analyze the physical consequences of  black  remnants, which form due to non-perturbative string theoretical  effects. These non-perturbative effects occur due to the T-duality in string theory. We will analyze the production of such  black remnants in models with large extra dimensions, and demonstrate that these non-perturbative effects  can explain the absence of mini black holes at the LHC.  In fact,  will  constraint such models using data from the LHC. We will  also analyze  such non-perturbative  corrections for various other black hole solutions. Thus, we will analyze the effects of such non-perturbative effects  on the Van der Waals behavior of AdS black holes. We will also discuss the effects of adding a chemical potential to this system. Finally, we will comment on the physical consequences of such  non-pertubative corrections  to black hole  solutions.
\end{abstract}

\maketitle
\section{Introduction}
It is known that black holes have an entropy, and this suggests that the physics of black holes can be described by some     quantum gravitational microstates.
It has been demonstrated that these microstates could be described using string theory \cite{mi12, mi16, mi18,  mi14}. So, it is both important and interesting to study the  black holes in string theory. In fact, it is known that black holes evaporate due to  the  Hawking radiation \cite{ther}. As the Hawking radiation is thermal, so it may not contain any information. Thus, there is the possibility of loss of information due to the evaporation of black holes, and this loss of information has led to  the information loss paradox \cite{ilpb}.
 It is hoped that quantum gravitational modification to the physical of black holes could help resolve such a paradox.  As string theory is one of the most well known approaches to quantum gravity,  it is hoped that the  black holes information paradox might get resolved in  string theory.  In fact, it has  been demonstrated that the thermodynamics  of black holes in string theory, can be obtained from the  microscopic states of the brane system   \cite{x1, x2, x3, x4}. Furthermore, these result   holds beyond the supergravity approximation.  \\
It has also been demonstrated that the folded strings, which are spontaneously created behind the horizon of the $SL(2,R)_k/U(1)$ black hole, can  violate the averaged null energy condition \cite{w1, w2}. The back-reaction from these folded string
can prevents information from falling into the black holes.  Furthermore, the
black hole entropy  can be related to  the number of these folded strings. The black hole information paradox in string theory has also been studied using the  diffeomorphism invariance for the Hilbert space \cite{w12}. This they seem to indicate that the black hole information paradox could be resolved in string theory. \\
It may be noted that    the black hole in string theory have also been studied using the   AdS/CFT correspondence \cite{w14, w16, w18, w20}. In fact, this has also led to the development of the fuzzball   proposal, in which the black holes are replaced by fuzzballs  \cite{w22, w24}.
As  fuzzballs do not have a horizon, the black holes are replacing  by objects without a horizon. However, it has been demonstrated that
 even though it is possible to reflect photons from the surface of a fuzzball,   the of such probability of such a reflection is very small \cite{w25}. So, the fuzzballs and the classical black holes
resemble each other, and fuzzballs can only be differentiated from classical black holes using certain  quantum effects.\\
It is also important to study the  black holes in string theory,   as they can have important cosmological consequences.
This is because the  physics of black holes in string theory has been used  to understand production of
 primordial black hole   \cite{w4, w5}.
These primordial black hole can be produced from an amplification of the power spectrum of curvature perturbations. This can in turn be achieved by    breaking the slow-roll conditions during inflation. As in the string axion models of inflation,   non-perturbative effects can  violate the slow-roll conditions, it is possible that such    string theoretical effects  can produce primordial black hole.
It is also possible to produce such an amplification from the
warp factor of D-brane with   a sufficiently large  step during the inflationary \cite{w5, w7}.  The production of primordial black holes has also been studied in brane world model \cite{b20, b22}.  In these brane world model, the physics of  primordial black holes is corrected  because of the modification in the Hawking temperature. Furthermore, in brane world models, there is an increase  the number of such   primordial black holes  in the  radiation-dominated epoch.  Thus, black holes in string theory  can have important  consequences for  important physical phenomena like the production of  primordial black hole. So, it is important to study the corrections to the physics of black hole from  string theory.\\
The entropy of a black hole solutions in heterotic string theory compactified on a torus has also been studied, and it has been demonstrated that this solution is characterized by a charge vector \cite{x12}.
Furthermore,  axion-dilaton black holes and Kaluza-Klein black holes are obtained for special values of the charge vector. It may be noted that the  entropy of a Kaluza-Klein black hole has also been obtained using the  Cardy formula \cite{x14}. Thus, string theory can lead to interesting consequences for black hole physics. \\
It is expected that the T-duality can lead to   ultraviolet finiteness of string theory \cite{s16, s18}, and this can in turn correct   black holes solutions  \cite{a, a1, a2}.  Furthermore, as the
double field theory is constructed using  T-duality \cite{df12, df14},
 the black hole thermodynamics in also invariant under the  duality of double field theory \cite{a4}.
 As such non-pertubative corrections modify the short distance
physics of black holes, they are expected to significantly modify the thermodynamics of the last stages of black holes evaporation   \cite{a}. It has been proposed that such modification to
a black hole would produce
a black remnant, and the entropy and heat capacity for  such a black remnant have been analyzed \cite{a}.
\\
It may be noted that such modifications of black holes  can only be obtained   from  non-perturbative effects,
and cannot be produced by pertubative calculations.  This is because T-duality can be used to demonstrate that the description of string theory below a certain zero point length is identical to its description above that zero point  length \cite{s16, s18}. It is thus possible to use T-duality to obtain a modified Green's function,  with such a  zero point length \cite{green1, green01, green02, green2}. This  Green's function modified by T-duality,  can be used to obtain the corrections to a  black hole solution  \cite{a}. Now this Green's function modified by T-duality,  does not diverge but   each term (in a pertubative series) tend to diverge   \cite{green1, green01, green02,  green2}. So, these corrections  cannot be obtained from specific terms in the perturbative series, and are only produced from non-perturbative string theoretical effects.
\\
It has been demonstrated that such corrections can produce a black remnant in a four dimensional Schwarzschild black hole \cite{a}. It would be interesting to generalization this results to higher dimensions.
This is because it is  expected that the black remnants would have interesting consequences for the detection of mini black holes at the LHC \cite{r2, r4}. This is because it is expected in models based on large extra dimensions, the effective Planck scale would be reduced in four dimensions \cite{P1, P2, P4, P5}. This occurs as in these models,  the standard model particles are localized on a brane, and the gravity is free to move in the bulk. These models were initially proposed as a   potential solution for the Hierarchy problem \cite{P1, P2, P4, P5}.  However, an interesting prediction of these models is the production of mini black holes  at the LHC \cite{bh12, bh14, bh16, bh18, Das, r6}. As such mini black holes have not been detected at the LHC  \cite{LHC, LHC3}, so there seems to be a problem with these models based on large extra dimensions. \\
However, it has also been suggested that if   black remnants form due to some non-perturbative effects, then these large extra dimensional models can still be valid \cite{r2, r4}. This is because black holes smaller than such black remnants cannot form or be detected at the LHC, and this would increase the energy needed to form such black remnants \cite{2r, 4r}. It may be noted that this has been used to explain the  absence  of such mini black holes at LHC \cite{r2, r4}.  Now as it has been observed that these  black remnants can also  form due to non-perturbative effects due to the   T-duality
\cite{a}, such  non-pertubative effects could explain the absence of detection of  mini black holes at  the LHC.
\\
So, in this paper, we will first analyze the corrections  to higher dimensional black holes from T-duality, and observe how they can produce black remnants in higher dimensions.  Then we will argue that such black remnants can
explain the absence of  mini black holes at the LHC, in models based on large extra dimensions. We will also analyze such corrections for various other black hole solutions, and comment on the physical consequences of such non-perturbative corrections.

\section{Black Holes Modified from T-Duality}
It   had been suggested that such black remnants should occur due to stringy theoretical effects \cite{r1}. This is because four dimensional magnetic black hole solutions of heterotic string theory can be constructed using $SU(2)/Z(2Q+2)$ WZW orbifold. It was observed that certain  marginal operators can deform this theory to an asymptotically flat black hole.  It was also argued  using the renormalization group flow that the gravitational singularity can be  resolved in this system, and black remnants can form in it. Thus, black hole remnants can form string theoretical effects. It is also possible to construct a regular solution for a  black hole due to T-duality \cite{a}.
This is because  the ultraviolet finiteness of string theory can modify the short distance physics
of black holes.
It may be noted that the  ultraviolet finiteness of string occurs  as the  geometry in string theory cannot  be probed   below a certain limit \cite{z2,zasaqsw}.
This is because in perturbative string theory, string length scale is the smallest probe available
\cite{cscds,2z}.    Even though non-perturbative   objects like  $D0$-branes also occur in string theory, it has been  argued that   space-time cannot be probed below $  \alpha'  g_s^{1/3}$, even with such non-perturbative objects
(where $g_s$ is the string coupling constant) \cite{s16, s18}.
This behavior of string theory occurs due to T-duality. As due to  T-duality, it is possible to show that the
description of string theory below the length $l_s = \alpha' $  is  the same as its description above $l_s = \alpha' $. \\
It may be noted that the bosonic string exists in
$26$ dimensions, so  the physics of  four  dimensions   can be obtained by compactifying
$22$ dimensions.   However, the important consequences of such compactification can also be understood by
analyzing  compactification with one compact dimension.
Now for a single compact dimension with radius $R$, the boundary conditions for a string can be written as \cite{s16}
\begin{equation}
 X^4 (\tau, \sigma + 2 \pi) = X^4 (\tau, \sigma)
 + 2 \pi w R,
\end{equation}
where $w$ is the winding number. The winding states for a string  are topologically
stable, and  will exist in the internal manifold with a non-contractible loop.
Now the   mass spectrum for such a system, can be written as
\begin{equation}
m^2 =\frac{1}{2\alpha'}\left(\, n^2 \frac{\alpha'}{R^2} +
w^2\frac{R^2}{\alpha'}\, \right)+ \dots,
\end{equation}
where $n$ is the Kaluza-Klein excitation level.
Now the important property of this spectrum is that  if we exchange  winding number $w$ and Kaluza-Klein excitation level $n$, this expression does not change. So, this expression is invariant under T-duality,
\begin{eqnarray}
w \to n, && R \to \alpha^{'2}/R.
\end{eqnarray}
This is the main reason for the description of string theory below the a certain length  being equivalent to its description above it.
As the double field theory is constructed using  T-duality \cite{df12, df14},
such a short distance behavior for  double field theory has also been studied   \cite{mi15}.  \\
The T-duality for the effective path integral of strings propagating in compactified extra-dimensions
has also been studied  \cite{green1, green01, green02, green2}.
This has been done by using   the string center of mass for analyzing the   Green's function for this system. So,  for such a  system, using  the string center of mass five-momentum $P_M\equiv \left(\, p_\mu\ , p_4\,\right)$,     propagation kernel can be written as
\begin{eqnarray}
&&K\left[\, x-y\ ; T\,\right]=\nonumber\\
&&\sum_{n=-\infty}^\infty
\int_{z(0)=x}^{z(T)=y}
\int_{x^4(0)=0}^{x^4(T)=n\,l_0}\left[\,Dz\,\right] \left[\,Dp\,\right]
\left[\, Dx^4\,\right]
 \left[\,Dp_4\,\right]\times\nonumber\\
&&\exp\left[\, i\int_0^T d\tau\left(\,p_\mu\dot x^\mu +p_4\dot x^4
-\frac{i}{2\mu_0}\left(\, p_\mu p^\mu +  p_4 p^4 \,\right)\,\right)\,\right]
\nonumber\\
&& \label{kernel}
\end{eqnarray}
where $\mu_0$ is  a  dimensional parameter. The contribution from the compact dimension are integrated to obtain the effective four dimensional propagator.
Here for a closed string (winding   around
the compact dimension)  all the oscillator
modes are neglected, and dynamics is expressed using its center of mass.
Now as two points are connected in the
 four dimensional space-time, the string  winds  around the compactified dimension.
So, the Green's function is obtained by taking a double sum over $n$ and $w$
\cite{green1, green2}.
It has been  demonstrated that this Green's function  is  invariant under T-duality,
$  w \to n,  R \to \alpha^{'2}/R$, with suitable identification of its parameters.
In momentum space, this  Green's function can be expressed  as   \cite{green1, green2},
 \begin{equation}
  G(k) = -\frac{2 \pi R }{\sqrt{k^2}} K_1 ( 2 \pi R \sqrt{k^2}),
 \end{equation}
where $K_1 (k)$ is a modified Bessel function of second kind. In this system a  zero point  length ($l = 2 \pi R$) is produced due to the compactified extra dimension of radius $R$, and the system cannot be probed below this length.
In the limit, $  2 \pi R k^2 = l k^2 \to 0$, the usual Green's function is obtained, $G(k) = - k^{-2}$.
This is the limit, in which the string is approximated as a particle, and the stringy effects are totally neglected.
This modified Green's function occurs due to smearing of the matter density from a
   point-like source to a distribution, due to the extended nature of strings.  This smearing of the matter density in turn modifies the geometry of a black hole. So, a corrected black hole solution is obtained by using the modified   matter density, which produces this modified Green's function. The metric for  such a four dimensional  static and spherically
symmetric black hole  modified by these string theoretical corrections  can be written as    \cite{a},
\begin{equation}\label{metric1}
ds^2=-f(r)dt^2+\frac{dr^2}{f(r)}+r^2 d\Omega^2,
\end{equation}
where $d\Omega^2=\sin^{2}\theta d\phi^{2}+d\theta^{2}$, and
\begin{equation}\label{sol1}
f(r)=1-\frac{2Mr^2}{(r^2+4\pi^{2}R^{2})^{3/2}}.
\end{equation}
Here $M$ denotes the Komar mass. This is a non-pertubative modification of a   static and spherically
symmetric black hole, and occurs due to the non-pertubative string theoretical effects. Such modifications of black holes  could not be obtained from   perturbative effects. This is because these corrections are obtained using a zero point length  corrected    Green's function, which does not diverge. However, this corrected Green's function  can be expressed as a sum of terms in a pertubative series, and  each of those terms would diverge \cite{green1, green01, green02, green2}.  Thus, these results cannot be obtained from specific terms in the perturbative series, and are only obtained from non-perturbative  string theoretical corrections. It may be noted that such string theoretical  corrections to black holes from  T-duality can non-trivially modify the behavior of black holes  \cite{a, a1, a2, a4}.  So, it is important to study the physical consequences of such modifications. In fact,  such non-pertubative corrections modify the short distance physics of black holes, they are expected to significantly modify the thermodynamics at the last stages of black holes \cite{a}.

\section{Black Remnant}
\begin{figure}[h!]
 \begin{center}$
 \begin{array}{cccc}
\includegraphics[width=70 mm]{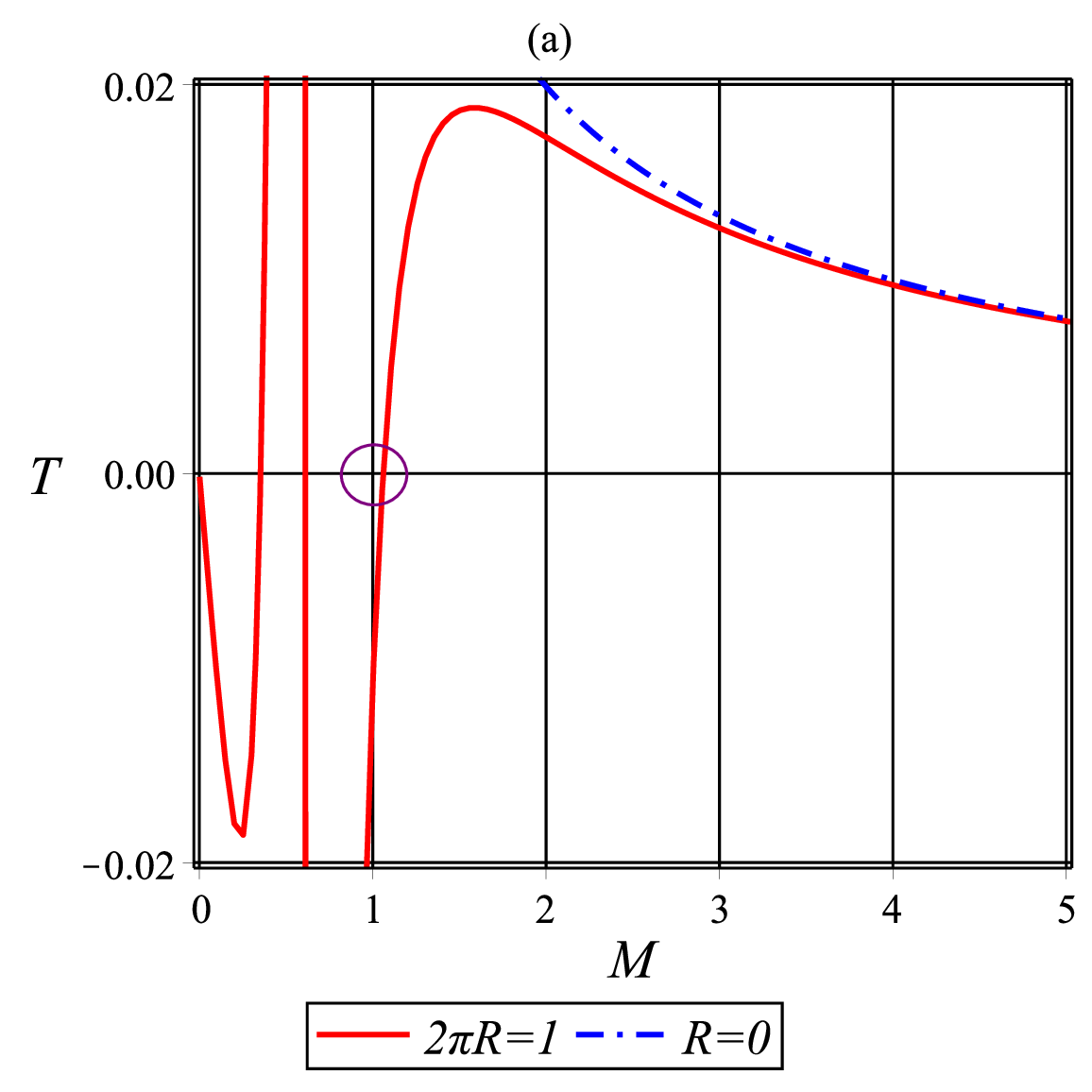}\includegraphics[width=70 mm]{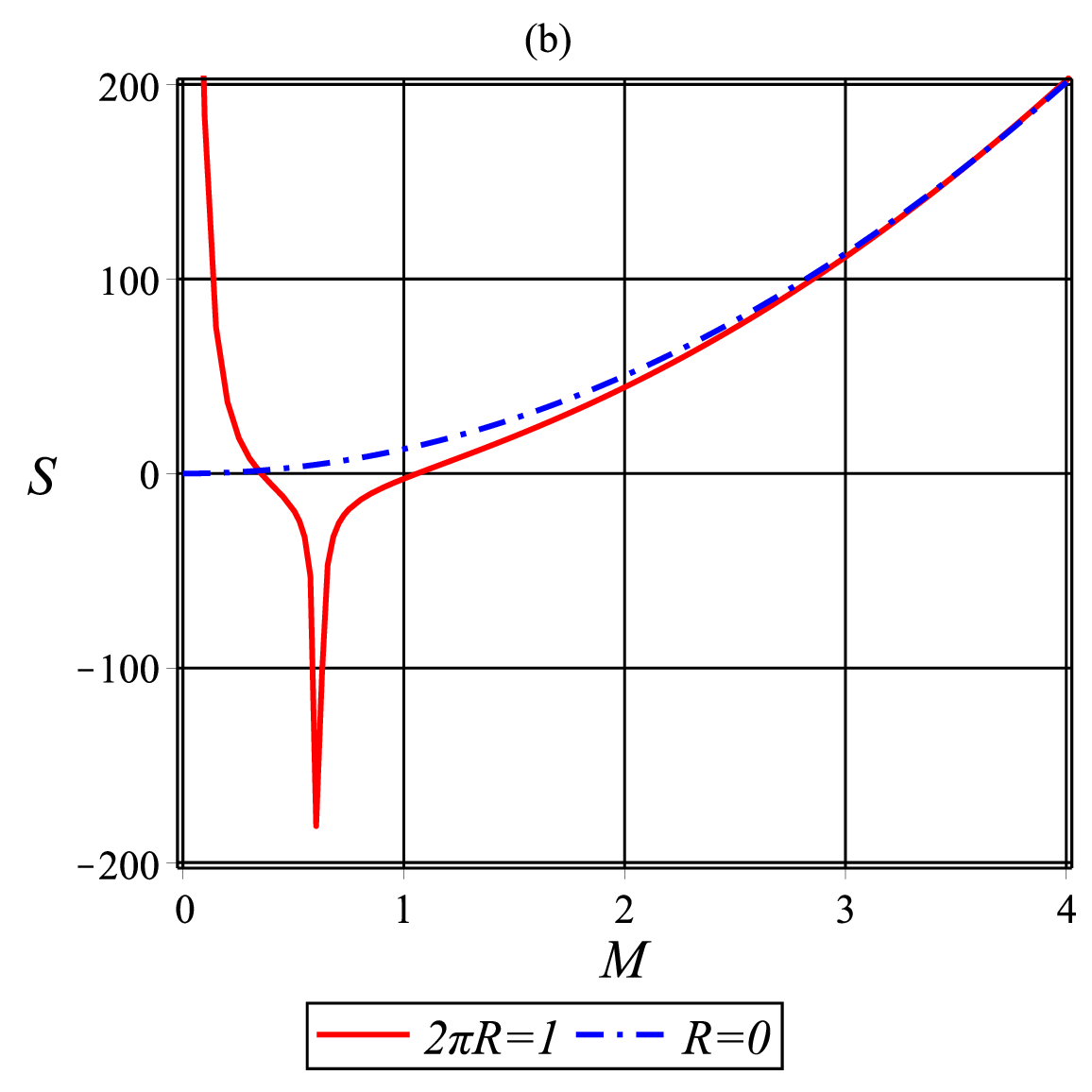}
 \end{array}$
 \end{center}
\caption{(a) Hawking temperature in terms of $M$. (b) Entropy in terms of $M$.}
 \label{1}
\end{figure}

It has  been observed that a Schwarzschild metric for a black hole changes to a Bardeen metric \cite{bardeen, bardeen1}, when the modifications due to T-duality are considered  \cite{a}.
Furthermore, as it is known that Bardeen black holes have a state where there temperature is zero, it has been observed that such states is a  black remnants \cite{bardeen, bardeen1}. So, a black remnant can form due to  T-duality in a four dimensional static spherical symmetric black hole  \cite{a}.
This is because, for such a corrected solution, and for a large mass ($M>3\sqrt{3}\pi^{2}R^{2}$), there exists two horizons \cite{a},
\begin{equation}
r_+\sim 2M-\frac{12\pi^{2}R^{2}}{4M},~~~ r_-\sim\frac{2\pi R}{\sqrt 2}\Big( \frac{2\pi R}{M}  \Big)^{1/2},
\end{equation}
where higher order ${R}/{M}$ terms  are neglected.
It is clear that the outer horizon produces  the Schwarzschild event horizon for the large mass. However, exact solution of $f(r)=0$ obtained for $M<3\sqrt{3}\pi^{2}R^{2}$, produces the following  event horizon radius,
\begin{equation}\label{h}
r_+=\frac{\sqrt{48M^{4}+16\alpha M^{3}+(12M^{2}-36\pi^{2}R^{2})\alpha^{2}-432M^{2}\pi^{2}R^{2}+\alpha^{4}+24\pi R M^{2}\sqrt{324\pi^{2}R^{2}-48M^{2}}}}{3\alpha},
\end{equation}
where $\alpha$ is given by
\begin{equation}
\alpha=(-108M\pi^{2}R^{2}+8M^{3}+6M\pi R\sqrt{324\pi^{2}R^{2}-48M^{2}})^{\frac{1}{3}}.
\end{equation}
Now as  $R\rightarrow0$, we obtain  $r_{+}=2M$.  However, as this is an imaginary solution, there is a lower bound for the black hole mass $M_e$, and only for $M>M_{e}={3\sqrt{3}\pi R}/{2}$, the black hole horizon exists \cite{Ivan3}.\\
As the T-duality modifies the Schwarzschild metric for a black hole changes to a Bardeen metric, and the Hawking temperature for a Bardeen black hole is known  \cite{bardeen, bardeen1}, it is possible to write the
  Hawking temperature of the stringy corrected black hole at the outer horizon as \cite{a},
\begin{equation}\label{9}
T=\frac{1}{4\pi r_+}\Big( 1-\frac{12\pi^{2}R^{2}}{r_+^2+4\pi^{2}R^{2}}  \Big),
\end{equation}
We should note that regions of interest for stringy T-dual black holes are for small mass or volume i.e.,  within the interval $[0,1]$.
It can be observed that   when $M$ is small, we have
\begin{equation}
 T \propto \sin{\kappa},
\end{equation}
where $\kappa$ is surface gravity.
However,   when $M$ is large, we have $T=c_{1}\kappa$. This unphysical behavior in the temperature for states smaller than a black remnant indicate that the state of zero temperature is actually a black remnant.
This is because the temperature behaves unphysically for black holes smaller than a black remnant \cite{a}.
In fact, from Fig. \ref{1} (a) suggests that temperature of black hole oscillates for mass $0<M<1$, i.e. increasing, decreasing,  and even vanishing  at certain point.
Even though it becomes non-zero beyond this point, physically the black hole cannot shrink beyond this point.
This is because when the temperature of the black hole becomes zero, it will stop radiating Hawking radiation, and form a black remnants \cite{a}. It is possible to calculate the mass of this black remnant.

\begin{figure}[h!]
 \begin{center}$
 \begin{array}{cccc}
\includegraphics[width=70 mm]{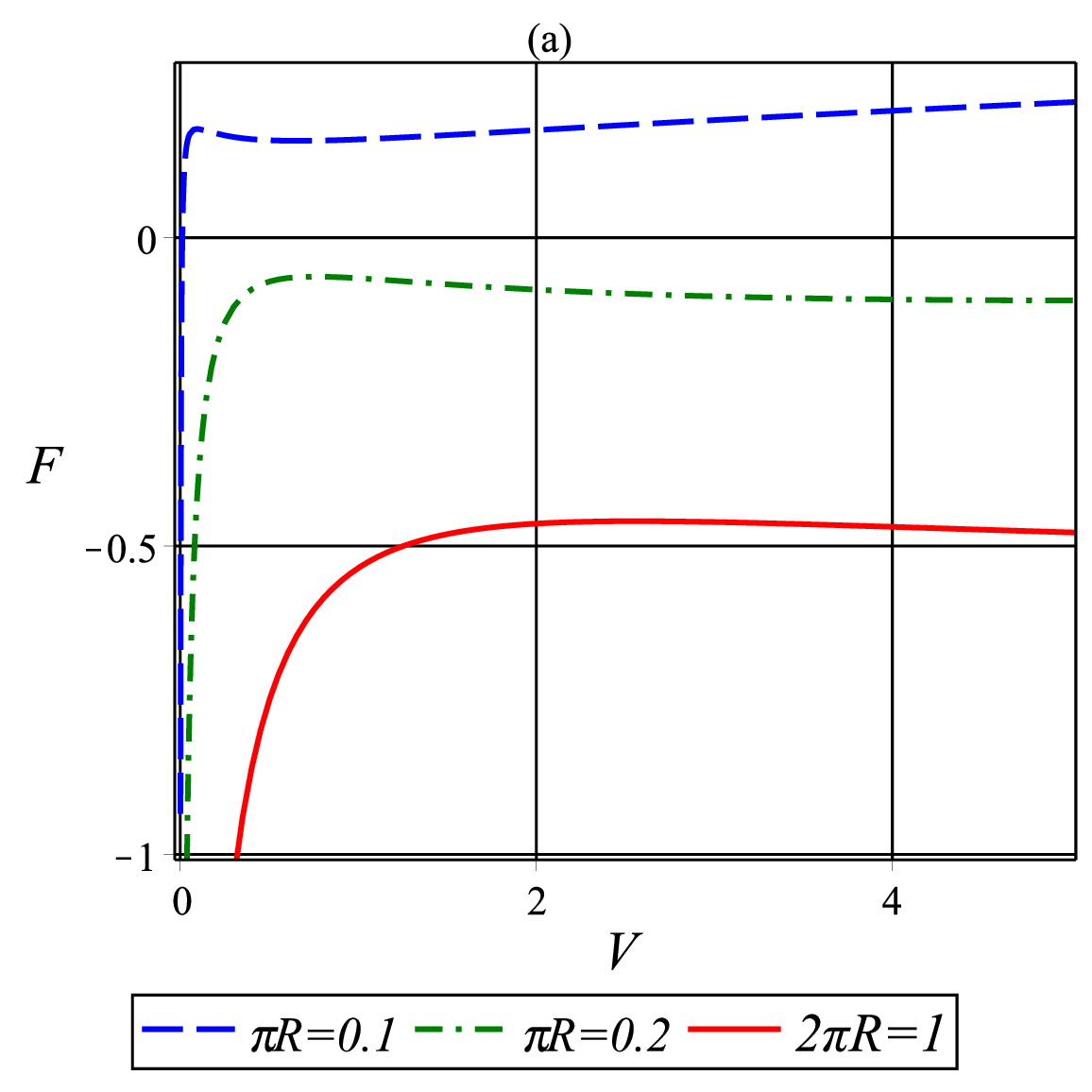}\includegraphics[width=70 mm]{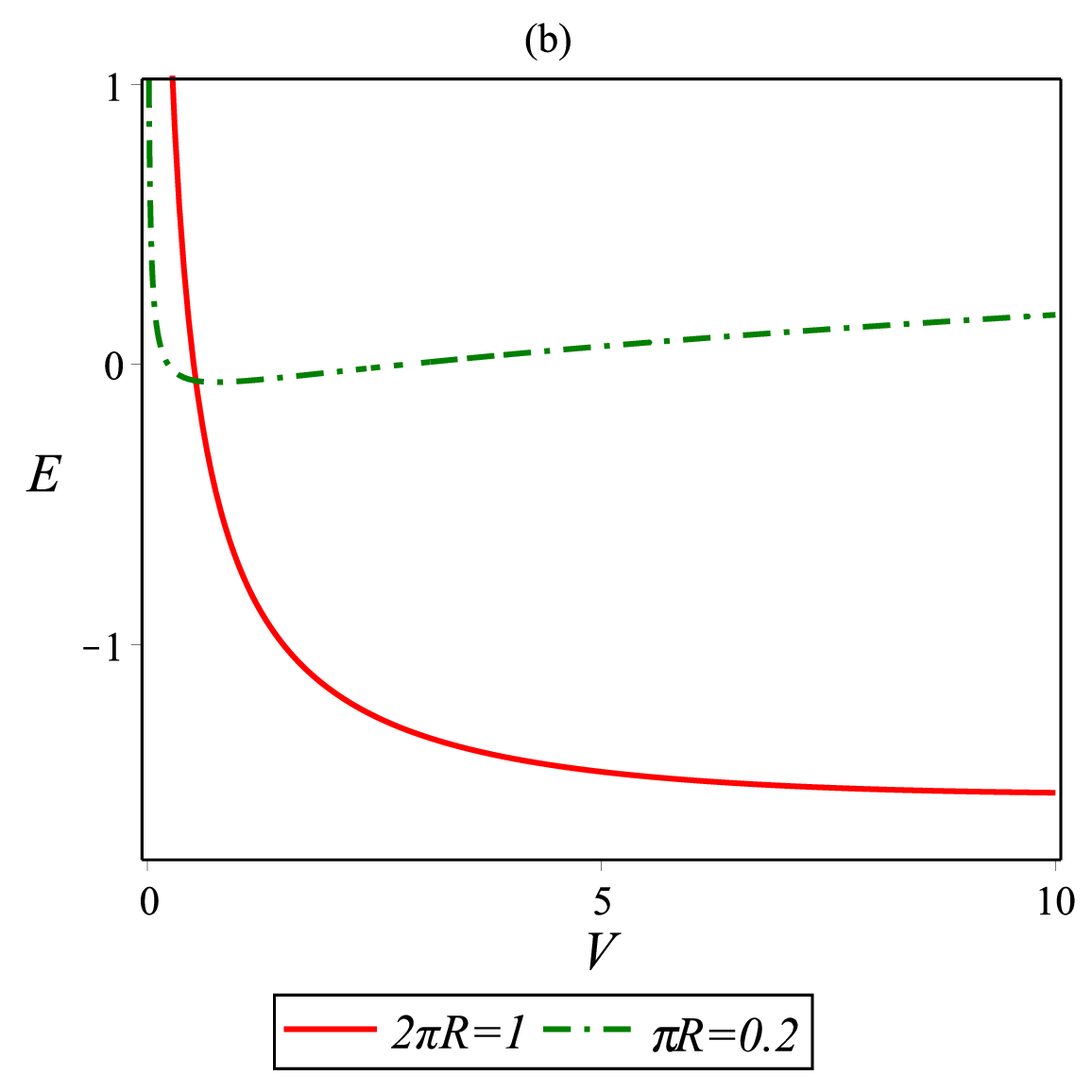}
 \end{array}$
 \end{center}
\caption{(a) Helmholtz free energy in terms of $V$. (b) Internal energy in terms of $V$.}
 \label{3}
\end{figure}

\begin{figure}[h!]
 \begin{center}$
 \begin{array}{cccc}
\includegraphics[width=80 mm]{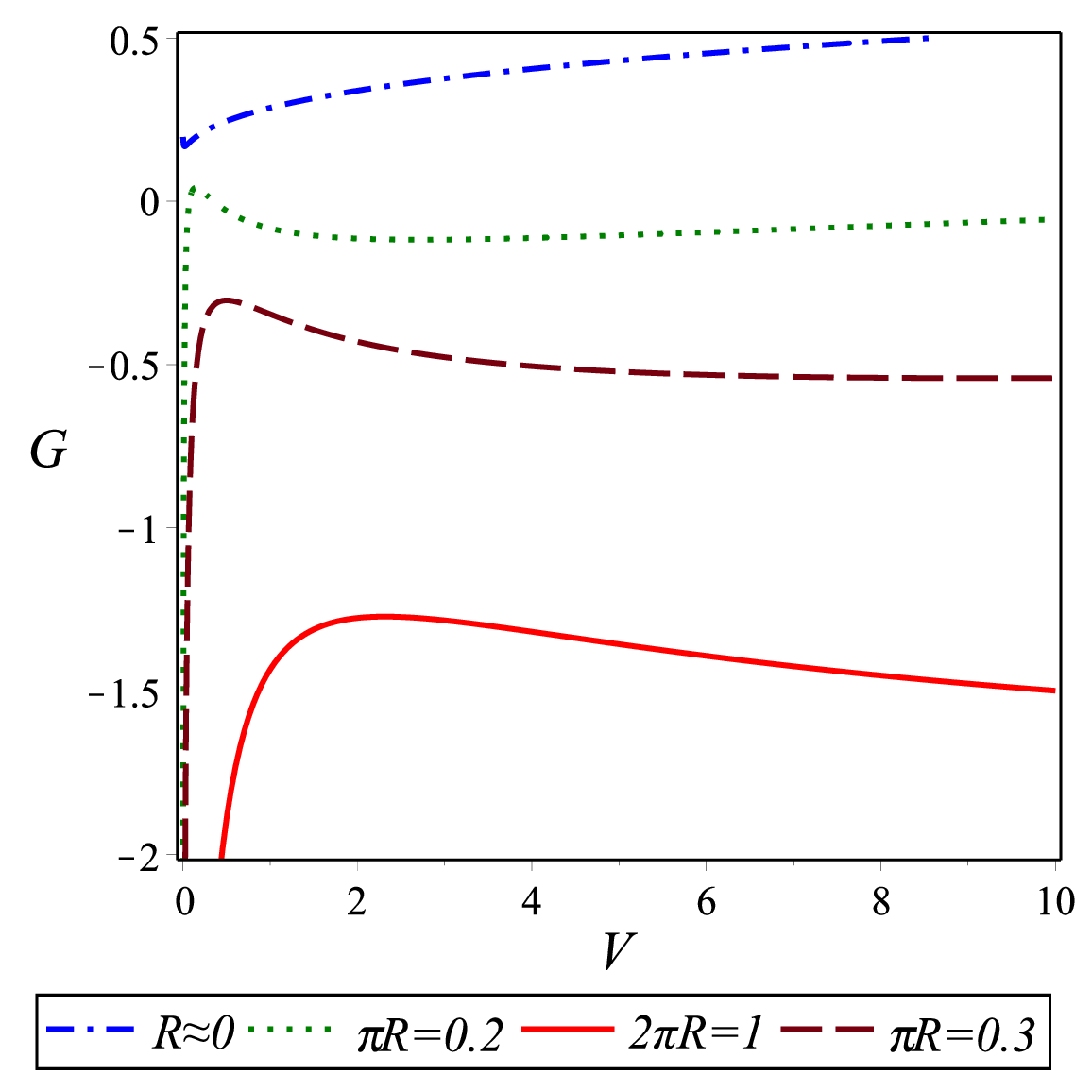}
 \end{array}$
 \end{center}
\caption{Gibbs free energy in terms of $V$.}
 \label{7}
\end{figure}

From the Fig. \ref{1} (a) we can see that there is a maximum mass ($M_max$), which is root of the following equation,
\begin{equation}
-32768{M}^{10}+94208{M}^{8}{(2\pi R)}^{2}-26624{M}^{6}{(2\pi R)}^{4}-9984{M}
^{4}{(2\pi R)}^{6}+4968{M}^{2}{(2\pi R)}^{8}-243{(2\pi R)}^{10}=0.
\end{equation}
For the $2\pi R=1$, we obtain  $M_{max}\approx1.5$.\\
Now because of  black hole evaporation, the black hole mass may decreased to $M_{min}$, where the black hole temperature is zero. It is possible to obtain an approximation value of this mass using  the violet circle in Fig. \ref{1} (a). However,  we can also obtain exact value by using Eqs.  (\ref{h}) and (\ref{9}). It is exactly the point where entropy become positive (Fig. \ref{1} (b)). Thus, from Eq.  (\ref{9}),  we can obtain $T=0$,  if $r_{min +}=2\sqrt{2}\pi R$, and so we can write \cite{a},
\begin{equation}
M_{min}\equiv M(T=0)=\frac{3\sqrt{3}}{2}\pi R.
\end{equation}
It may be noted that it is known that Bardeen black holes have such a state, where the temperature is zero \cite{bardeen, bardeen1}. In fact, as T-duality changes the Schwarzschild metric to a Bardeen metric, it has been argued that the black hole temperature would become zero because of T-duality at a certain critical size \cite{a}.
This is the minimum mass of the black remnant from due to string theory corrections.
\\
It is also important to analyze other thermodynamic quantities for such a black hole corrected from these string theory corrections.
Since $A=4\pi r_+^2$ denotes the area of the horizon and $V=\frac{4}{3}\pi r_+^3$ is the volume of the black hole, the black hole entropy given by \cite{a},
\begin{eqnarray}\label{corrected entroy}
S&=&\frac{4(M^{4}-5 \pi^{2} R^{2}M^{2}+\frac{9\pi^{4}R^{4}}{4})}{M^{2}}\sqrt{1+\frac{\pi^{2}R^{2}}{(2M-\frac{12\pi^{2}R^{2}}{4M})^{2}}}\nonumber\\
&-&12\pi^{3}R^{2}\left(\sinh^{-1}\sqrt{2}-\frac{1}{\sinh^{-1}(\frac{2\pi R}{2M-\frac{12\pi^{2}R^{2}}{4M}})}\right).
\end{eqnarray}
In the Fig. \ref{1} (b), we can analyze the  behavior of entropy. The entropy  is an increasing function of $M$, which becomes equal  to the Schwarzschild entropy for the large $M$. This is expected as the string theoretical effects should only modify the short distance behavior of the black holes. So, in the infrared limit, the black hole resembles the usual black hole obtained from general relativity. In the Fig. \ref{1} (b),
 the entropy has a sharp minimum ($M\approx0.5$) and than sudden increase as $M$ further increases.
This behavior is completely different from  classical black hole thermodynamics, and
$S\propto A$ does not hold with $0<M<1$. In fact, entropy and area have a different relation in this limit, which can be approximated to
\begin{equation}
 S = \frac{c_{1}}{(M-a)^n}
\end{equation}
where $a$ and $n$ are positive numbers while $c_{1}$ is an arbitrary constant. The it seems that there  is a power-law form of entropy in the  ultraviolet  limit. This indicates that the black holes smaller than remnants are not physical, due to the universality of area-entropy relation in black holes.
 \\
Helmholtz free energy for this stringy corrected black hole  can obtain by using the following relation,
\begin{equation}\label{Helmholtz}
F=-\int SdT.
\end{equation}
 We can obtain analytic expression of the free energy in terms of the black hole volume. In the Fig. \ref{3} (a), we analyze  the relation between the Helmholtz free energy and  volume $V$. We observe that  there is a maximum for value for it, for a certain value of  volume.
Internal energy for such black holes corrected from T-duality can be written as (with  $2\pi R=1$)
\begin{eqnarray}
E&=&F+ST
\nonumber \\
&\approx&\frac{1}{6}\left(\frac{1}{3}(1-5V^{-\frac{2}{3}})\xi-V^{-\frac{1}{3}}\right)\left(1-\frac{3}{\zeta^{2}}\right)\nonumber\\
&-&\frac{5\mu\left(100(7V^{2}+20)\tan^{-1}(2\nu)+(3V^{2}+7)\tanh^{-1}(\frac{100}{\nu})\right)}{\nu(4V^{\frac{2}{3}}+10)},
\end{eqnarray}
where $\xi$ and $\zeta$ are given by
\begin{eqnarray}
\xi=\sqrt{9V^{\frac{2}{3}}+24}, &&
\zeta=\sqrt{0.4V^{\frac{2}{3}}+1}.
\end{eqnarray}
In order to analyze its exact behavior,  we analyze it numerically. Now using Fig. \ref{3} (b),  we can analyze the behavior of the internal energy. There is critical volume with special value of $l$, where internal energy is minimum. As $l  = 2\pi R$, so by varying the value of the compactification radius the internal energy of this black hole solution can also change.\\
\\
It is important to understand the behavior of specific hear for this system, and analyze the string theoretical corrections to the specific hear of this black hole.
We can write  the specific heat in terms of $M$ as (for $l=2\pi R=1$) \cite{a}
\begin{eqnarray}
C&=&T\left(\frac{dS}{dT}\right) \nonumber \\ &\approx& - {\frac {4\times 10^{8}}{\sqrt {{\delta}^{2}-1} \left( 8{\mathcal{M}}^{2}
- 3 \right)^{4}{\mathcal{M}} \left(  4\times 10^{3}{\mathcal{M}}^{8}- 10^{4}{\mathcal{M}}^{6}+
 8\times 10^{3}{\mathcal{M}}^{4}- 2\times10^{3}{\mathcal{M}}^{2}+ 81 \right) }},
\end{eqnarray}
where $\delta=\sqrt { 1+ \left(  2M- 0.75{M}^{-1} \right)^{-2}}$ and
$\mathcal{M}= {M}^{20}-3.5{M}^{18}+5.4{M}^{16}+5{M}^{14}+ 3{M}^{12}- 1.4{M}^{10}
+ 0.5{M}^{8}- 0.1{M}^{6}+ 0.015{M}^{4}- 0.0014{M}^{2}+ 0.000055. $
\begin{figure}[h!]
 \begin{center}$
 \begin{array}{cccc}
\includegraphics[width=80 mm]{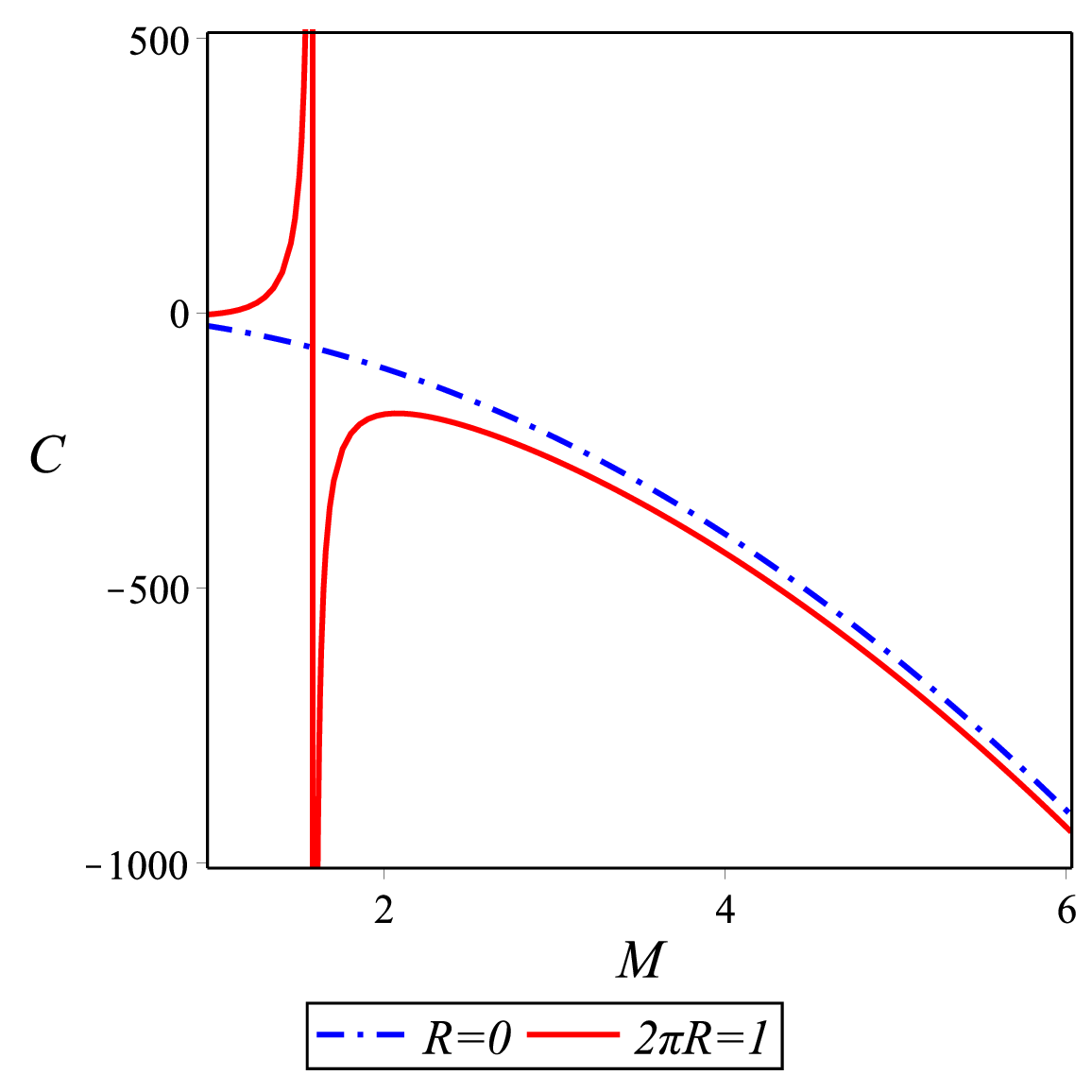}
 \end{array}$
 \end{center}
\caption{Specific heat in terms of $M$.}
 \label{5}
\end{figure}
It has been argued  that   there is the first order phase transition around the black remnant,  due to the divergence of the    specific heat  \cite{a}.
We plot behavior of the specific heat  \cite{a}, in terms of the black hole mass in the Fig. \ref{5} and show that in the limit $R\rightarrow0$ the black hole is completely unstable, while there is some stable regions when $R\neq0$. These regions are corresponding to the lower mass of black hole. Now an unstable phase for black hole is obtained by increasing  the mass of the  black hole. Using  the Fig. \ref{5},  we can show that there is a discontinuity within range $0<M<2$. The specific heat goes from initially
positive to negative values showing a second order phase transition. It confirms that black holes with small mass and string correction are stable while classical large Schwarzschild like black holes are unstable. Furthermore, for small black hole masses $ M=M_{min}$, the specific heat goes to zero at $C=0$. This indicates the existence of remnants,  because when the
heat capacity is zero, the black hole cannot exchange radiation with the surrounding space \cite{gupf}. \\
In order to investigate stability of the system, we calculate Gibbs free energy
\begin{equation}
 G=H-TS
\end{equation}
The minimum of Gibbs free energy  represent   stability, and its maximum represent unstable to stable phase transition (corresponding to asymptotic
behavior of the Fig. \ref{5}). The Gibbs energy will always be decreasing for a  system in equilibrium. From the Fig. \ref{7},  we can observe  that the Schwarzschild black hole is unstable. However, the string theoretical corrections to the Schwarzchild black hole can make it stable for small black hole radius. This is an important change in the physical behavior of this system, which occurs due to short distance corrections coming from string theory.

\section{Higher dimensions}
It is expected that in models with large extra dimensions, the effective Planck scale would be reduced in four dimensions, and this can lead to a potential solution for the Hierarchy problem \cite{P1, P2, P4, P5}.  In such models, the lowering of effective Planck scale should have lead to the production of mini black holes at the LHC \cite{bh12, bh14, bh16, bh18, Das, r6}. As such mini black holes have not been detected at the LHC  \cite{LHC, LHC3},  this was initially taken as an indication for the absence of such large  extra dimensions.
However, it has also been suggested that   black remnants can have some important implication for the detection
of black holes at the LHC \cite{r2, r4}. This is because  if black remnants forms, then black holes smaller than such black remnants cannot form or be detected at the LHC.  So, such black remnants would  modify the minimum size of  such black holes which could be produced at the LHC (which is the size of remnants) \cite{2r, 4r}. In fact, this can be used as a reason for the absence  of such black holes at LHC \cite{r2, r4}.  As black remnants  form due to non-perturbative effects
\cite{a}, these non-pertubative effects could explain the absence of  such mini black holes  at the LHC. So, we will will demonstrated  that such black remnants can also form due to a modification of higher dimensional black hole solution by  T-duality. We will  also argue that mini black holes at the LHC have not  been detected  due to the  such black remnants, which form because  of T-duality of string theory.
To analyze the production of such black remnants in these models, we need to study
 the thermodynamics of black holes modified by T-duality in higher dimensions.

\begin{figure}[h!]
 \begin{center}$
 \begin{array}{cccc}
\includegraphics[width=60 mm]{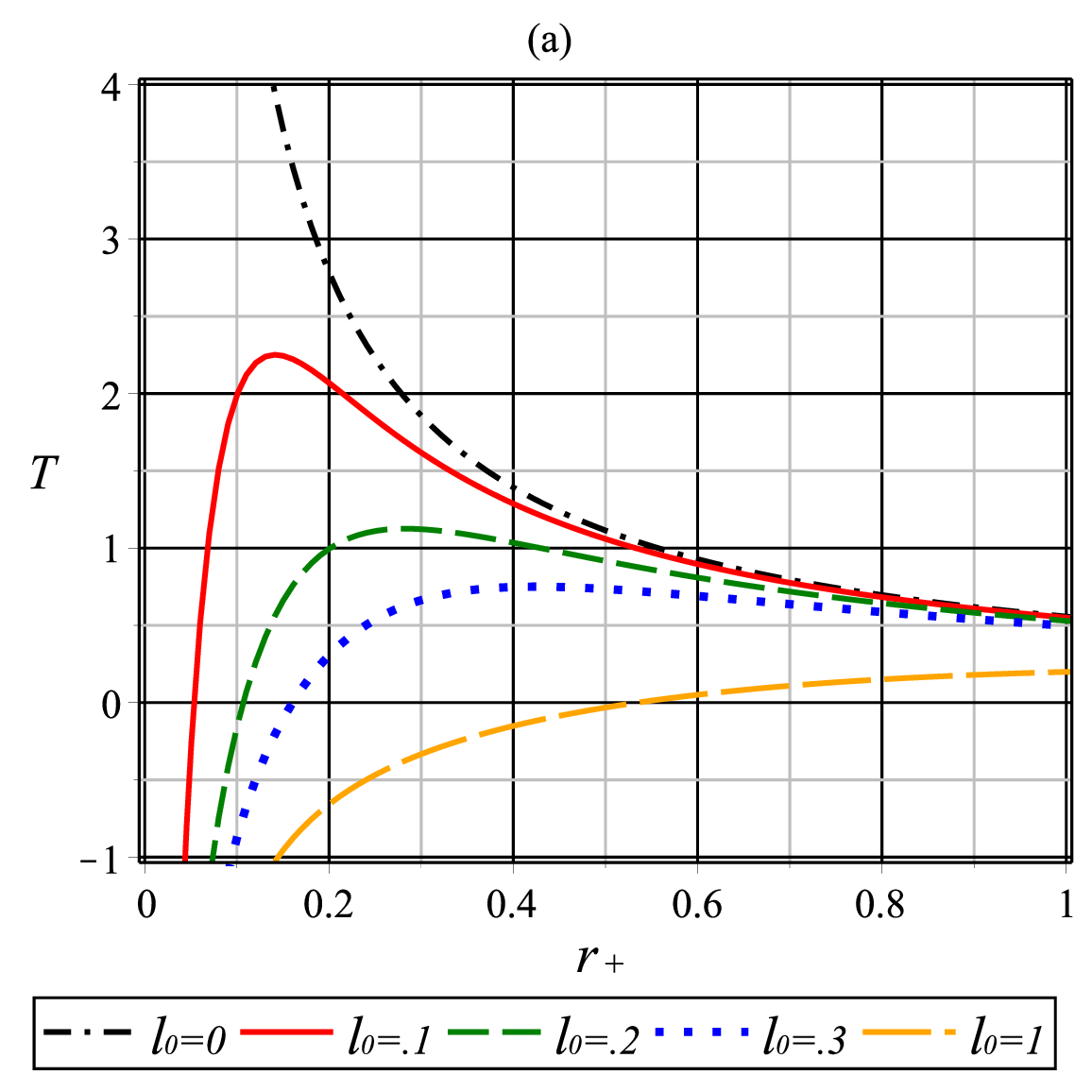}\includegraphics[width=60 mm]{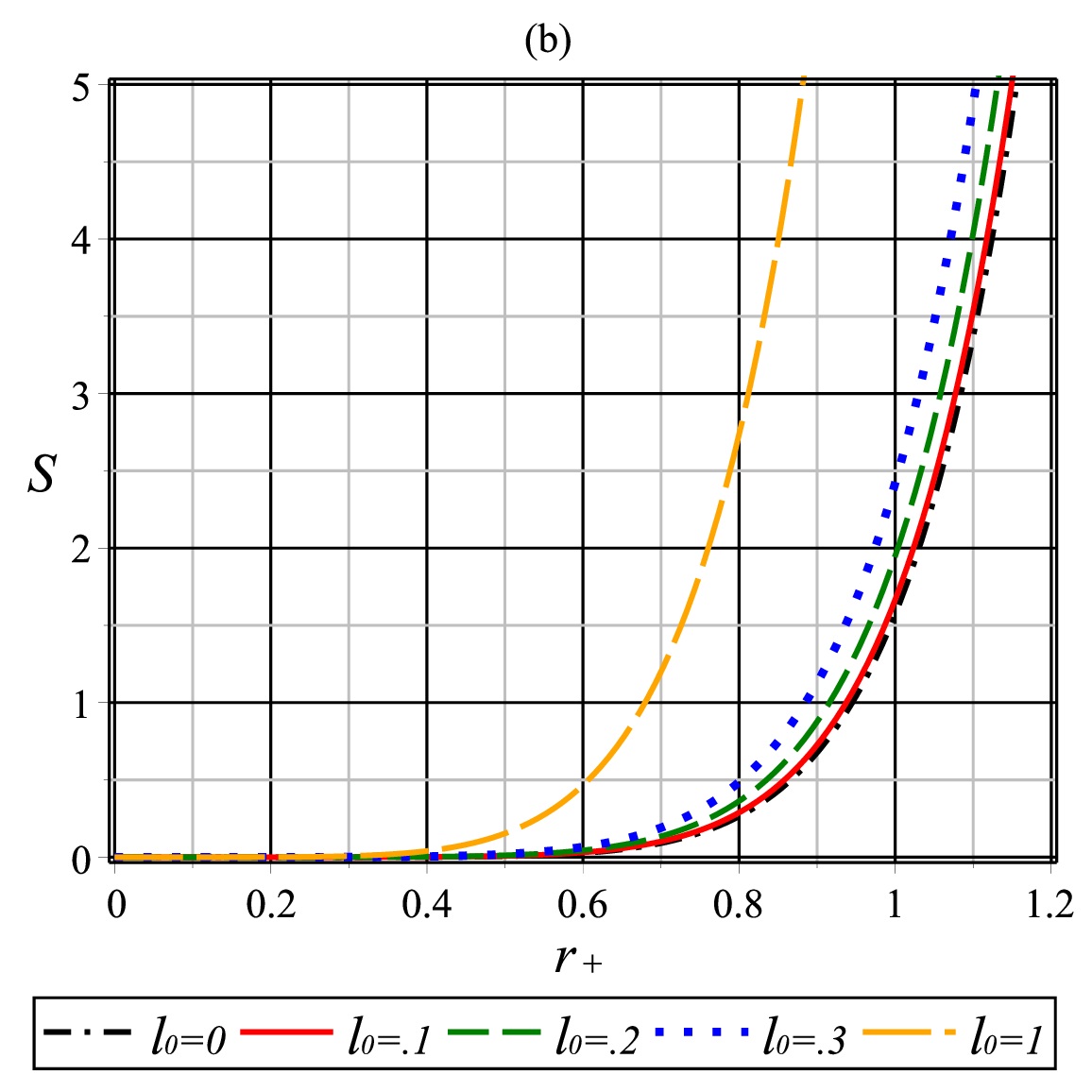}\includegraphics[width=60 mm]{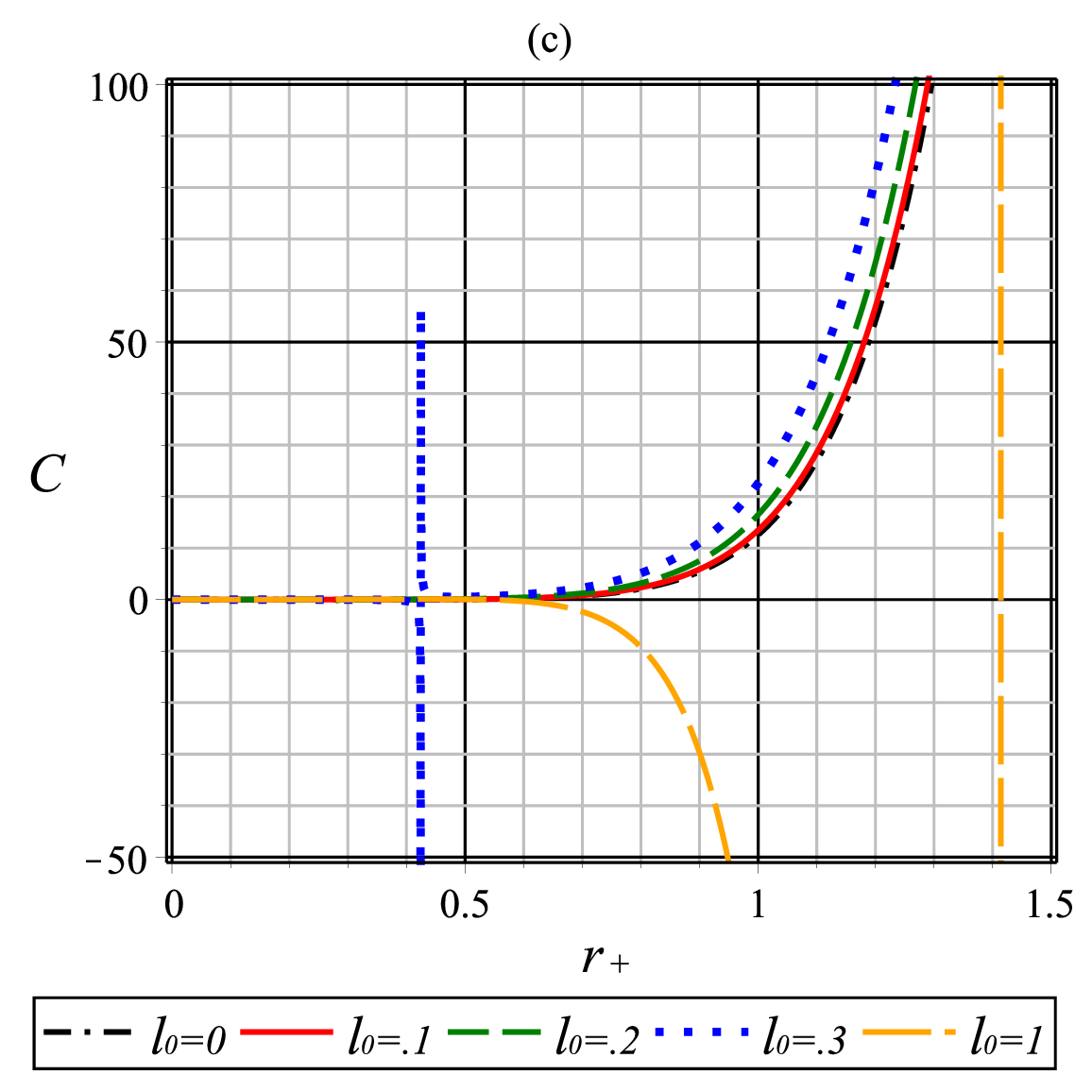}
 \end{array}$
 \end{center}
\caption{(a) Temperature (b) entropy and (c) specific heat in terms of $r_{+}$ for the case of $d=10$ in units of $b$.}
 \label{11}
\end{figure}

It is possible to generalize the results obtained from T-duality \cite{a},
to higher dimensional black holes. This is because it has been argued \cite{green1, green01, green02, green2}, that
 a toroidal compactification  can be used to  obtain a Green's function with zero point length,
 in higher dimensions. The zero point length $l_0$  depend on the geometry of the compactified manifold. Thus, it would depend on the radius of   compact dimensions.   In the simple case, of a single compactified dimensions, with the $R$ as the  radius compactification,   $l_0 = 2\pi R$. In extra dimensions, the geometric dependence of such a zero point length would be more complicated, however, T-duality would relate it to  a unique zero point length in the theory \cite{green1, green01, green02, green2}. This new modified Green's function could be obtained by summing over the various Kaluza-Klein and winding modes. Thus it is possible to obtain corrections to a higher dimensional black hole by using this Green's function modified by a zero point length.  This corrected black hole solution  should have an intrinsic zero point length in the metric. We also observe that such a modified higher dimensional black hole, should reduce to a usual higher dimensional Schwarzchild black hole at large distances. Thus, these corrections have to be scale dependent.  As the angular part of the black hole does not have an explicit scale, it would not be modified by such corrections. Furthermore, we expect that this modification should coincide with the corrected four dimensional black hole \cite{a}, when $d =4$. Thus, we can write the metric for a black hole in   higher dimensional black hole, corrected by T-duality as
\begin{equation}\label{metric22}
ds^2=-f(r)dt^2+\frac{dr^2}{f(r)}+r^2 d\Omega_{d-2}^2,
\end{equation}
where $d\Omega_{d-2}^2$ is the metric for the $(d-2)$ dimensional  unite sphere (this is not modified as we expect these corrections to be scale dependent). Now  the modification only  occurs in $f(r)$,  as
\begin{equation}\label{hi1}
f(r)=1-\frac{\mu r^2}{(r^2+l_{0})^{\frac{d-1}{2}}},
\end{equation}
where    $\mu$ is related to the black hole mass as \cite{hi12, hi14}
\begin{equation}\label{hi1-1}
\mu=\frac{16\pi G_{d}M}{(d-2)\Omega_{d-2}}\equiv\frac{M}{b},
\end{equation}
with  $b$ as a constant related to Planck mass. It may be noted that for $d =4$, this metric  coincides with the four dimensional black hole corrected by T-duality \cite{a}.  It also reduces to a higher dimensional Schwarzchild black hole at large distances, when the effects from zero point length can be neglected. However, at short distances, the metric is has an intrinsic zero point length.
It may be noted that the Planck mass in $d$ dimensions can be written using the  $d$ dimensional Newton's constant $G_{d}$ as
\begin{equation}
 M_{P}^{d-2}=G_{d}^{-1}
\end{equation}
\\
Now we can write the  temperature of this higher dimensional black hole as
\begin{equation}\label{hi2}
T=\frac{[(d-3)r_{+}^{2}-2l_{0}^{2}]\mu r_{+}}{4\pi(r_{+}^2+l_{0}^{2})^{\frac{d+1}{2}}}=\frac{(d-3)r_{+}^{2}-2l_{0}^{2}}{4\pi r_{+}(r_{+}^2+l_{0}^{2})},
\end{equation}
where in the last equality,  we have  used
$\mu={(r_{+}^2+l_{0}^{2})^{\frac{d-1}{2}}}/{r_{+}^{2}}.
$ In the case of $l_{0}=0$,  we recover results of  \cite{r2, r4}.
The first law of thermodynamics can   be written  as,
\begin{equation}\label{hi4}
dM=TdS.
\end{equation}
Hence, the black hole entropy for black holes with  $d>4$ is  given by,
\begin{equation}\label{hi5}
S\approx\frac{2\pi b}{(d-4)(d-2)}(2(d-4)r_{+}^{2}+l_{0}^{2}(d-2)(d-1))r_{+}^{d-4}+S_{0},
\end{equation}
where $S_{0}$ is an integration constant,  and we have neglected $\mathcal{O}(l_{0}^4)$.
The constant $S_{0}$ can be fixed, using the fact that   at $T=0$ the entropy should vanish. Thus, we obtain
\begin{equation}\label{hi5-1}
S_{0}=-\frac{2\pi b l^{d-2}(\frac{2}{\sqrt{2d-6}})^{d-4}(d^{3}-6d^{2}+15d-22)}{(d-4)(d-3)(d-2)}.
\end{equation}
Now, we can obtain the specific heat for this black hole
\begin{equation}\label{hi6}
C=\frac{2(r_{+}^{2}+l_{0}^{2})((d-3)r_{+}^{2}-2l_{0}^{2})(2r_{+}^{2}+l_{0}^{2}(d-1))b\pi r_{+}^{d-4}}{(d-3)r_{+}^{4}-(d+3)l_{0}^{2}r_{+}^{2}-2l_{0}^{4}},
\end{equation}
In the plots of the Fig. \ref{11},  we can see behavior of thermodynamics quantities like temperature, entropy and specific heat for some values of $l_{0}$,  in ten dimensions.
In the case of $l_{0}=0$,  we can see that temperature  is increased by decreasing radius of the horizon  (see Fig. \ref{11} (a)). However,  in presence of stringy corrections, the temperature becomes  zero, and forms a    black remnant. The entropy of the black hole also becomes  zero at small scales  (see Fig. \ref{11} (b)). As illustrated by Fig. \ref{11} (c), the  specific heat is increasing function of horizon radius for the small $l_{0}$. So, by increasing value of $l_{0}$,    phase transition occurs.  \\
Now, it is clear that at $r_{min +}={2l_{0}}/{\sqrt{2d-6}}$,  we have
\begin{eqnarray}
 T (r_{min +})=0, & S (r_{min +}) =0, & C (r_{min +}) =0.
\end{eqnarray}
 It may be noted that in  the case of $d=4$, (with $l_0 = 2 \pi R$),
 the results for a four dimensional black hole \cite{a} are recovered.  \\
Now using $r_{min +}={2l_{0}}/{\sqrt{2d-6}}$,  we can obtain, the mass of the black remnant for a black hole
modified by T-duality as
\begin{eqnarray}\label{hi7}
M_{min}\equiv M(T=0)&=&\frac{b}{2}l_{0}^{d-3}\sqrt{\frac{(d-1)^{d-1}}{(d-3)^{d-3}}}\nonumber\\
&=&\frac{(d-2)\pi^{\frac{d-3}{2}}M_{P}^{d-2}}{16\Gamma(\frac{d-1}{2})}l_{0}^{d-3}\sqrt{\frac{(d-1)^{d-1}}{(d-3)^{d-3}}}.
\end{eqnarray}
In models with extra dimensions, mini black holes can form in four dimensions, due to the reduction of the Planck scale in four dimensions, in such models. Thus, in such models, mini black holes can form at the LHC.
In such models,  the   standard model particles are localized on a   brane, and gravity propagates freely in the higher dimensional bulk. In these models, it is possible to calculate the emission rate   from the temperature of a black hole using  the Stefan-Boltzmann law. Thus,    for a black hole on a brane, we can write $M_{t}\equiv  {dM}/{dt}=-a A T^{4},$  \cite{bh12, bh14, bh16, bh18, Das},
where $a$ is a positive constant depend on the effective Stefan-Boltzmann constant and Planck mass.
It is possible to write the total  total cross section for a collision that produces a black hole as \cite{4}
\begin{equation}\label{hi8}
\sigma\approx\pi r_{+}^{2}.
\end{equation}
Hence, the differential cross section for the production of a mini black hole modified by T-duality can be written as
can be written as
\begin{equation}\label{hi9}
\frac{d\sigma}{dM}=\frac{(d-2)\pi^{\frac{d-1}{2}}M_{P}^{d-2}r_{+}^{4}}{4(r_{+}^{2}+l_{0}^{2})^{\frac{d-3}{2}}\Gamma(\frac{d-1}{2})((d-3)r_{+}^{2}-2l_{0}^{2})}.
\end{equation}
Using LHC results \cite{LHC, LHC3},  we know that $M_{min}>13 TeV$. So, using  (\ref{hi7}),  we can write
\begin{equation}\label{hi10}
l_{0}>\left(\frac{208\Gamma(\frac{d-1}{2})}{(d-2)\pi^{\frac{d-1}{2}}M_{P}^{d-2}}\sqrt{\frac{(d-3)^{d-3}}{(d-1)^{d-1}}}\right)^{\frac{1}{d-3}}.
\end{equation}
For the lower bound, we obtain $l_{0}$ for some values of $M_{P}$ for various dimensions in Table 1. It is expected that    energies of about  $100 \, TeV$ would be reached  in near future colliders  \cite{LHC2, LHC4},
and so   we have also include   this energy in our analysis in Table 2.  This is the bound on zero point  length that can be detected at these energies. If such a bound exists, then the mini black holes would not be observed at energies below those  energies. Thus, it is possible that we have not observed the mini black holes at the LHC, due to these non-perturbative effects, as they  would increase the energies need for the production of such mini black holes.

\begin{center}
  \begin{tabular}{|@{} l @{} ||@{} c @{} | @{}r @{}|@{} c @{} |@{} c @{} |@{} c @{} ||@{} l @{} ||@{} c @{} | @{}r @{}|@{} c @{} |@{} c @{} |@{} c @{} |}
    \hline
    d           & $l_{0}>$     & $M_{P}$  &   \\ \hline\hline
    6           & 0.095        & 4.54     &   \\ \hline
    7           & 0.139        & 3.51     &   \\ \hline
    8           & 0.180        & 2.98     &   \\ \hline
    9           & 0.212        & 2.71     &   \\ \hline
    10          & 0.243        & 2.51     &   \\ \hline
    \hline
  \end{tabular}
\end{center}
Table 1. Mass of the black hole remnant, in different dimensions for the energy of order 13 TeV.

\begin{center}
  \begin{tabular}{|@{} l @{} ||@{} c @{} | @{}r @{}|@{} c @{} |@{} c @{} |@{} c @{} ||@{} l @{} ||@{} c @{} | @{}r @{}|@{} c @{} |@{} c @{} |@{} c @{} |}
    \hline
    d           & $l_{0}>$     & $M_{P}$  & \\ \hline\hline
    6           & 0.188        & 4.54     & \\ \hline
    7           & 0.231        & 3.51     & \\ \hline
    8           & 0.269        & 2.98     & \\ \hline
    9           & 0.866        & 2.71     & \\ \hline
    10          & 0.326        & 2.51     & \\ \hline
    \hline
  \end{tabular}
\end{center}
Table 2. Mass of the black hole remnant, in different dimensions for the energy of order 100 TeV.

\section{Other Black Hole Solutions}
\begin{figure}[h!]
 \begin{center}$
 \begin{array}{cccc}
\includegraphics[width=60 mm]{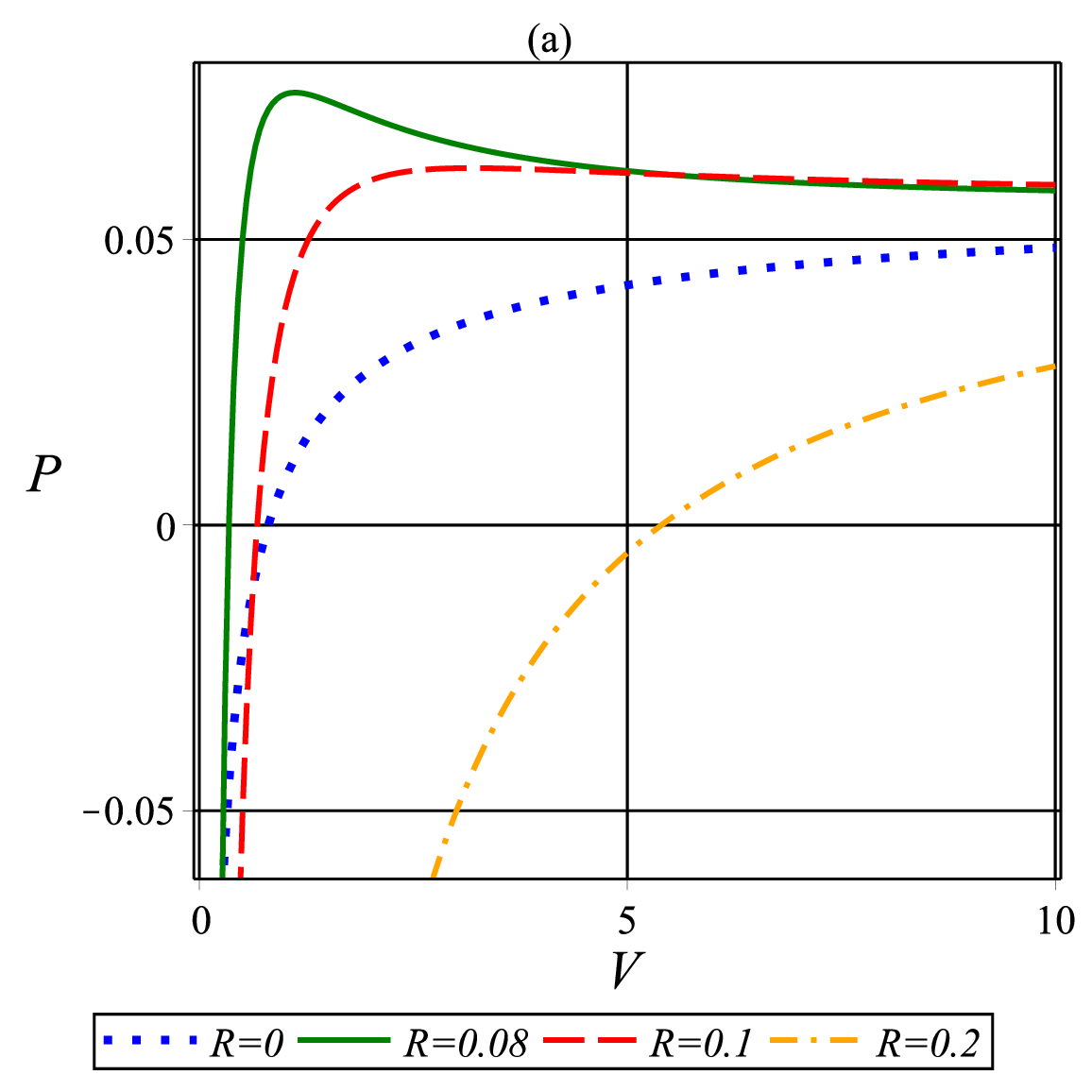}\includegraphics[width=60 mm]{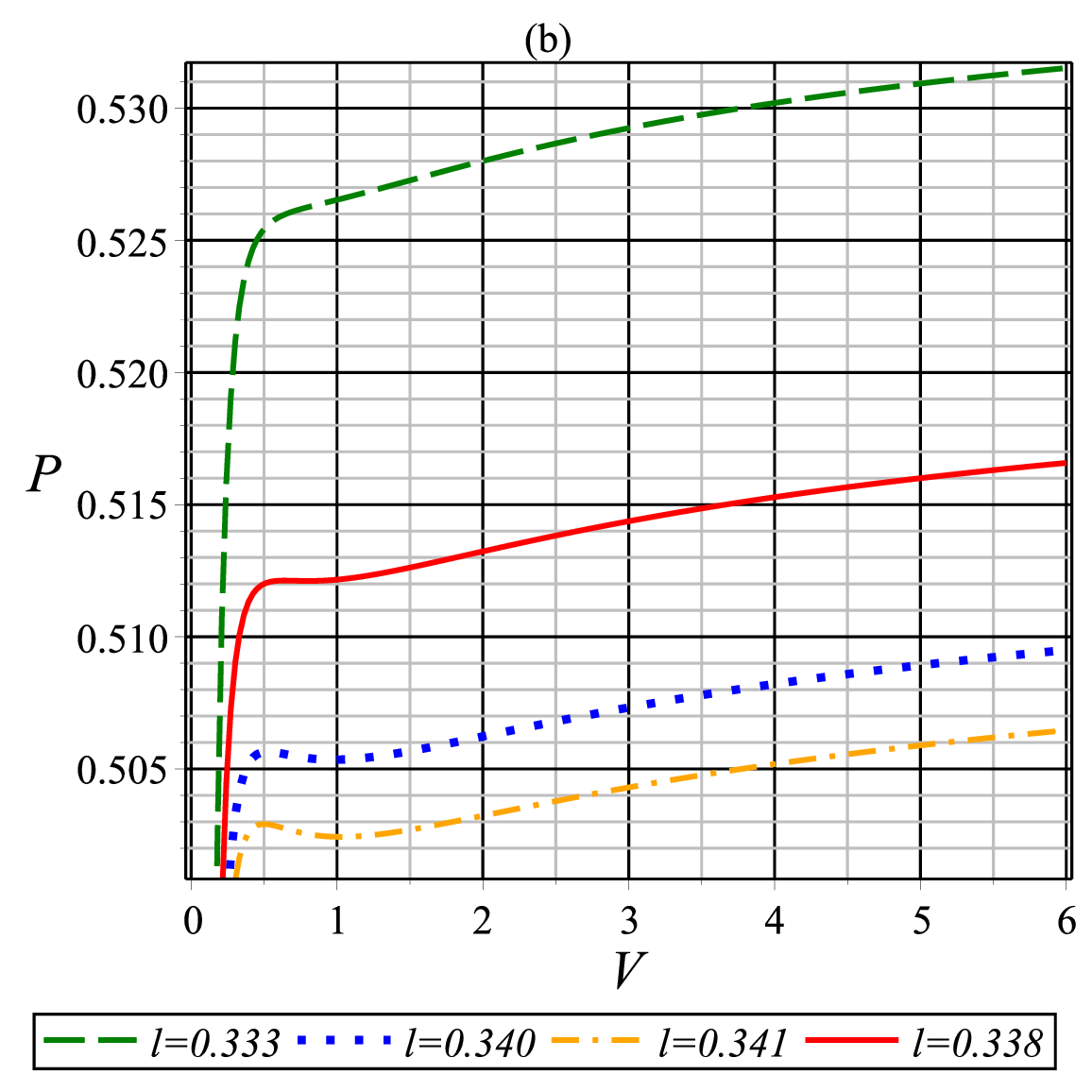}
 \end{array}$
 \end{center}
\caption{Pressure in terms of $V$. (a) $l=1$, (b) $R=0.04$.}
 \label{8}
\end{figure}

It is possible to obtain corrections to  other black hole solutions, and analyze the physical consequences of such a short distance modification on those solutions. This can be done by first noting that only the modification occurs due to extended structure of strings, and their  wrapping  in extra dimensions
\cite{green1, green01, green02, green2}. Thus, the  T-duality only modifies the form of matter density in a black hole solution \cite{a}, and this modification would not modify  other constants directly. However, due to the modification of matter part of the black hole solutions, non-trivial effects can occur for other black hole solutions. Thus, it also possible to analyze the modification to different  black holes from T-duality.
Thus, we can first analyze these effect for a modified AdS black hole.   This solution can be obtained by using the matter density and a cosmological constant term \cite{a}.
As the cosmological constant term occurs due to a vacuum expectation value of  fields, and we are only interested analyzing the modification to  a simple AdS black hole solution, we do not expect that the cosmological constant term will be modified by the T-duality. It might be possible that the T-duality could change the value of this  cosmological constant term, but these corrections to the cosmological constant can be absorbed into the new cosmological constant term, and we would can still write the corrected AdS solution, with only corrections to the matter fields. So, using the corrections to  the matter density \cite{a}, which is obtained from  the T-duality \cite{green1, green2}, we can write the metric for a modified AdS black hole  as
\begin{equation}\label{Cosmol1}
f(r)=1-\frac{2Mr^2}{(r^2+4\pi^{2}R^{2})^{3/2}}+\frac{r^{2}}{L^{2}},
\end{equation}
where $\Lambda\propto-L^{-2}$ is the  cosmological constant term and
\begin{equation}\label{Cosmol2}
M=\frac{(1+\frac{r_{+}^{2}}{L^{2}})(r_{+}^2+4\pi^{2}R^{2})^{3/2}}{2r_{+}^{2}},
\end{equation}
is the modified Komar mass. It may be noted that for $\Lambda =0$, it reduced to the  spherical symmetric   black hole modified by T-duality \cite{a}. Furthermore, at large distances, it can be approximated by a AdS
black hole. So, now we write the temperate for this black hole as
\begin{equation}\label{Cosmol3}
T=\frac{3r_{+}^{4}+L^{2}(r_{+}^{2}-8\pi^{2}R^{2})}{4\pi L^{2}r_{+}(r_{+}^{2}+4\pi^{2}R^{2})},
\end{equation}
which is reduced to the temperature given by Eq. (\ref{9}),  if $L\rightarrow\infty$ ($\Lambda=0$). It is possible to identify the pressure with
  cosmological constant as \cite{PV1, PV2, PV3, PV4}
\begin{equation}\label{Cosmol4}
P=-\frac{\Lambda}{16\pi}=\frac{L_{c}}{16\pi L^{2}},
\end{equation}
where $L_{c}$ is a constant.
Now it is possible to write the  first law of thermodynamics for a AdS black hole with pressure as the cosmological constant  \cite{PV1, PV2}. Even thought the specific values of the thermodynamic quantities  are modified by T-duality, we can still write the first law of thermodynamics as
\begin{equation}\label{first law}
dM=TdS+PdV.
\end{equation}
So,  using the  first order approximation,  it is possible to  write  entropy  (\ref{corrected entroy}) corrected by T-duality as
\begin{equation}\label{Cosmol4}
S=\pi r_{+}^{2}-3\pi^{3} R^{2}+ {{\mathcal{O}}(r_{+}^{-2})},
\end{equation}
where $r_{+}$ is largest real root of the following equation,
\begin{equation}\label{Cosmol5}
1-\frac{2Mr^2}{(r^2+4\pi^{2}R^{2})^{3/2}}+\frac{r^{2}}{L^{2}}=0.
\end{equation}
Furthermore, we can also now obtain other thermodynamic quantities for this black hole corrected by T-duality. So, by using the Eq.  (\ref{Helmholtz}),  we can obtain Helmholtz free energy for the AdS black holes corrected by T-duality as
\begin{eqnarray}\label{Cosmol6}
F&=&{\frac {9\pi^{2} R^{2}r_{+}}{4{L}^{2}}}-{\frac {3{\pi}^{2}R^{2}}{2r_{+}}}-{\frac {3{\pi }^{2}{R}^{2}r_{+}}{{L}^{2}}}-{\frac {{r_{+}}^{3}}{4{L}^{2
}}}+\frac{r_{+}}{4}-{\frac {12{\pi }^{4}{R}^{
4}r_{+}}{ \left( 4{\pi }^{2}{R}^{2}+{r_{+}}^{2} \right) {L}^{2}}}\nonumber\\
&-&{\frac {9{\pi }^{4}{R}^{4}r_{+}}{ \left( 4{\pi }^{2}{R}^{2}+{r_{+}}^{2}
 \right) {L}^{2}}}+{\frac {3{\pi }^{2}{R}^{2}r_{+}}{4{\pi }^{2}{R}^{2}
+{r_{+}}^{2}}}+{\frac {9\pi^{2} R^{2}r_{+}}{16{\pi }^{2}{R}^{2}+16{r_{+}}^{2}}}\nonumber\\
&+&{\frac {12{\pi }^{3}{R}^{3}}{{L}^{2}}\tan^{-1} \left({\frac {r_{+}}{2
\pi R}} \right) }-3\pi R\tan^{-1} \left({\frac {r_{+}}{2\pi R}}
 \right).
\end{eqnarray}
It may be noted that as in the limit in which we the cosmological constant  becomes zero, the corrected Helmholtz free energy reduced to the the Helmholtz free energy of a Schwarzschild black hole.
It is possible  to write it in terms of volume and obtain thermodynamics pressure,
\begin{equation}\label{Cosmol7}
P=-\left(\frac{\partial F}{\partial V}\right)_{T}.
\end{equation}
This  should be identical with pressure given by the Eq. (\ref{Cosmol4}). In the plots of Fig. \ref{8},
we observe the  behavior of pressure in terms of $V$. We can see from Fig. \ref{8} (a) that there is a minimum volume for the system, below that the pressure is negative,  and it is corresponding to phase transition, which can be  obtained from   Gibbs free energy. From solid red line of Fig. \ref{8} (b) it is clear that there is critical point when,
\begin{equation}\label{Cosmol8}
\left(\frac{\partial P}{\partial V}\right)_{T}=\left(\frac{\partial^{2} P}{\partial V^{2}}\right)_{T}=0.
\end{equation}

\begin{figure}[h!]
 \begin{center}$
 \begin{array}{cccc}
\includegraphics[width=70 mm]{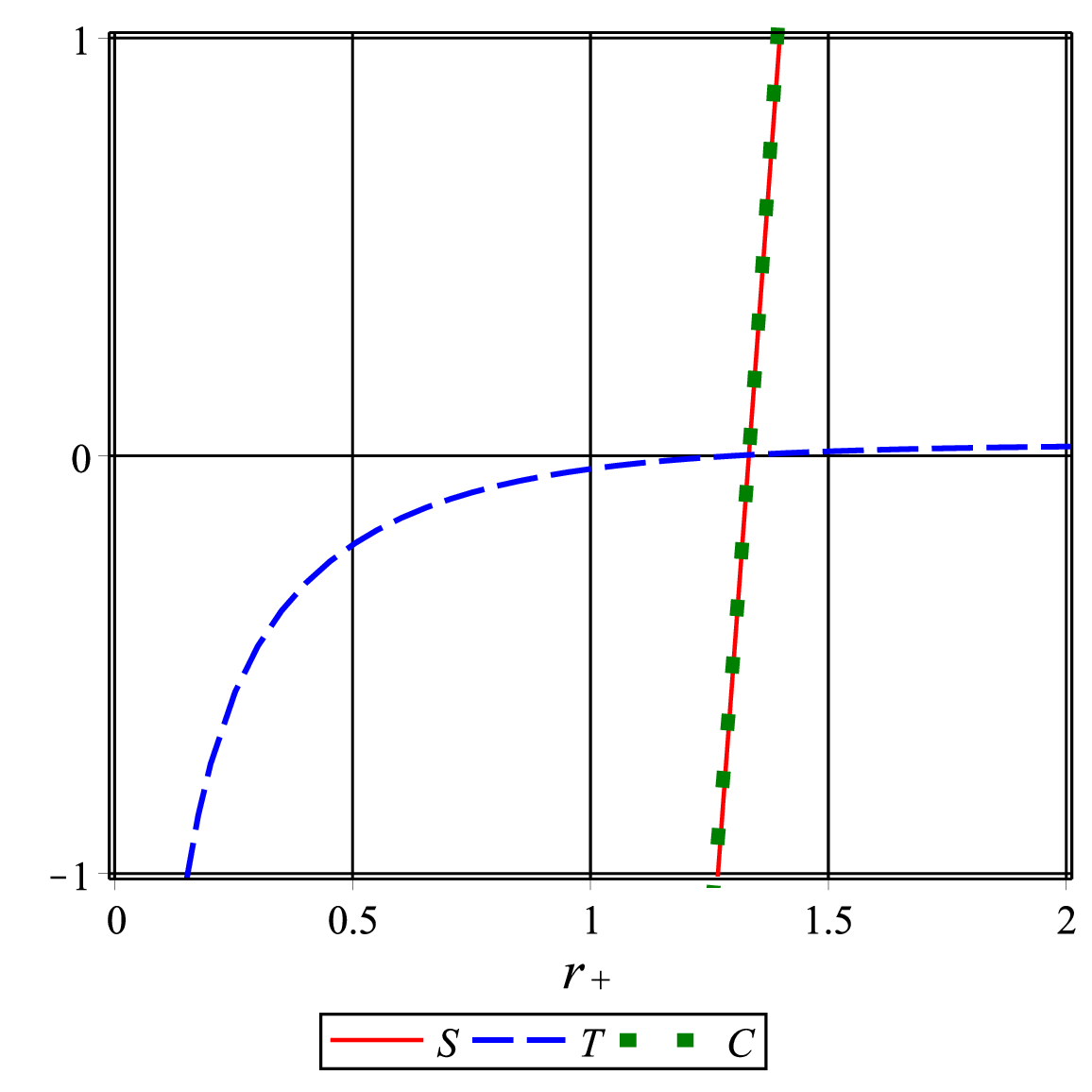}
 \end{array}$
 \end{center}
\caption{Temperature, entropy and specific heat in terms of $r_{+}$ for $L=10$ and $R=0.15$.}
 \label{88}
\end{figure}

\begin{figure}[h!]
 \begin{center}$
 \begin{array}{cccc}
\includegraphics[width=65 mm]{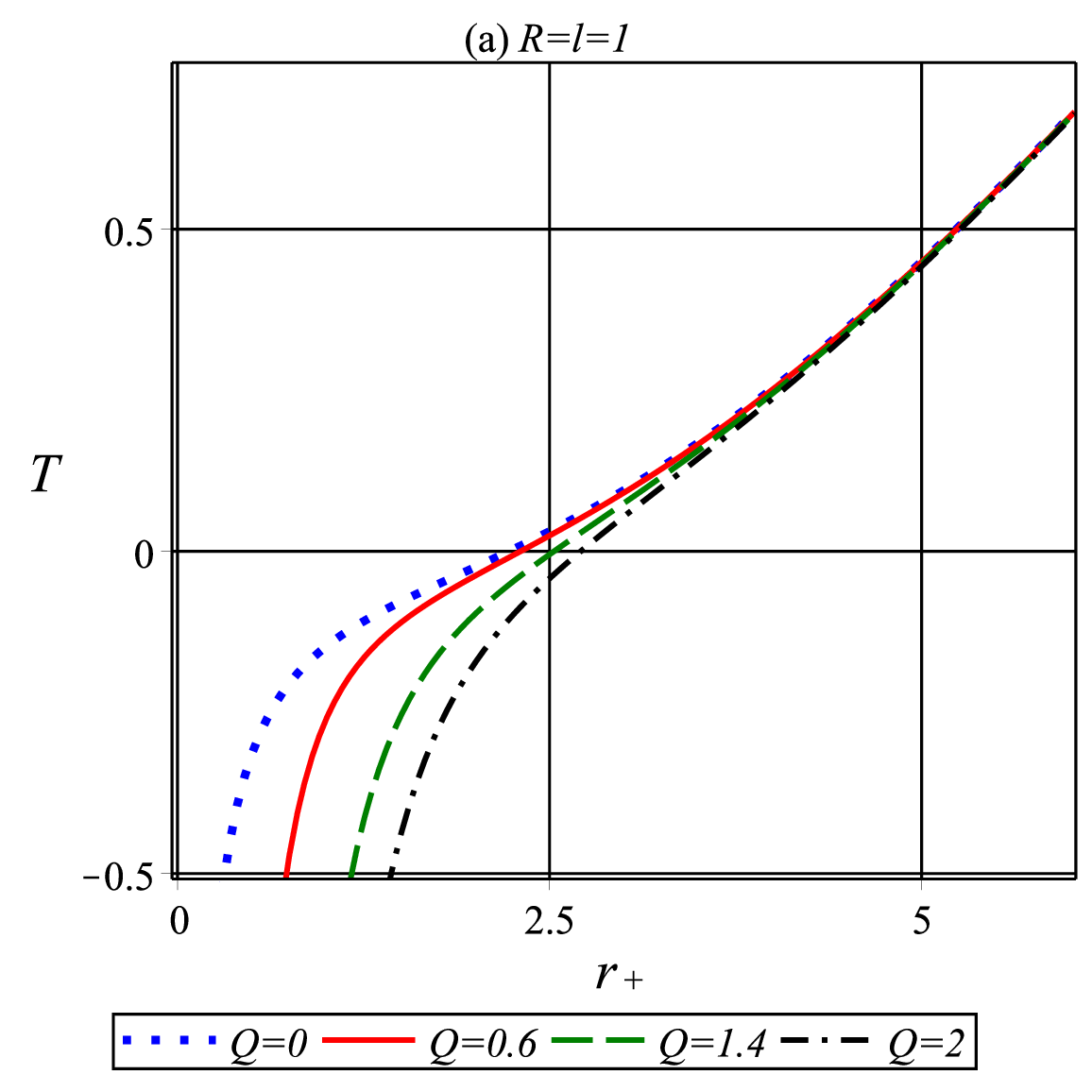}\includegraphics[width=65 mm]{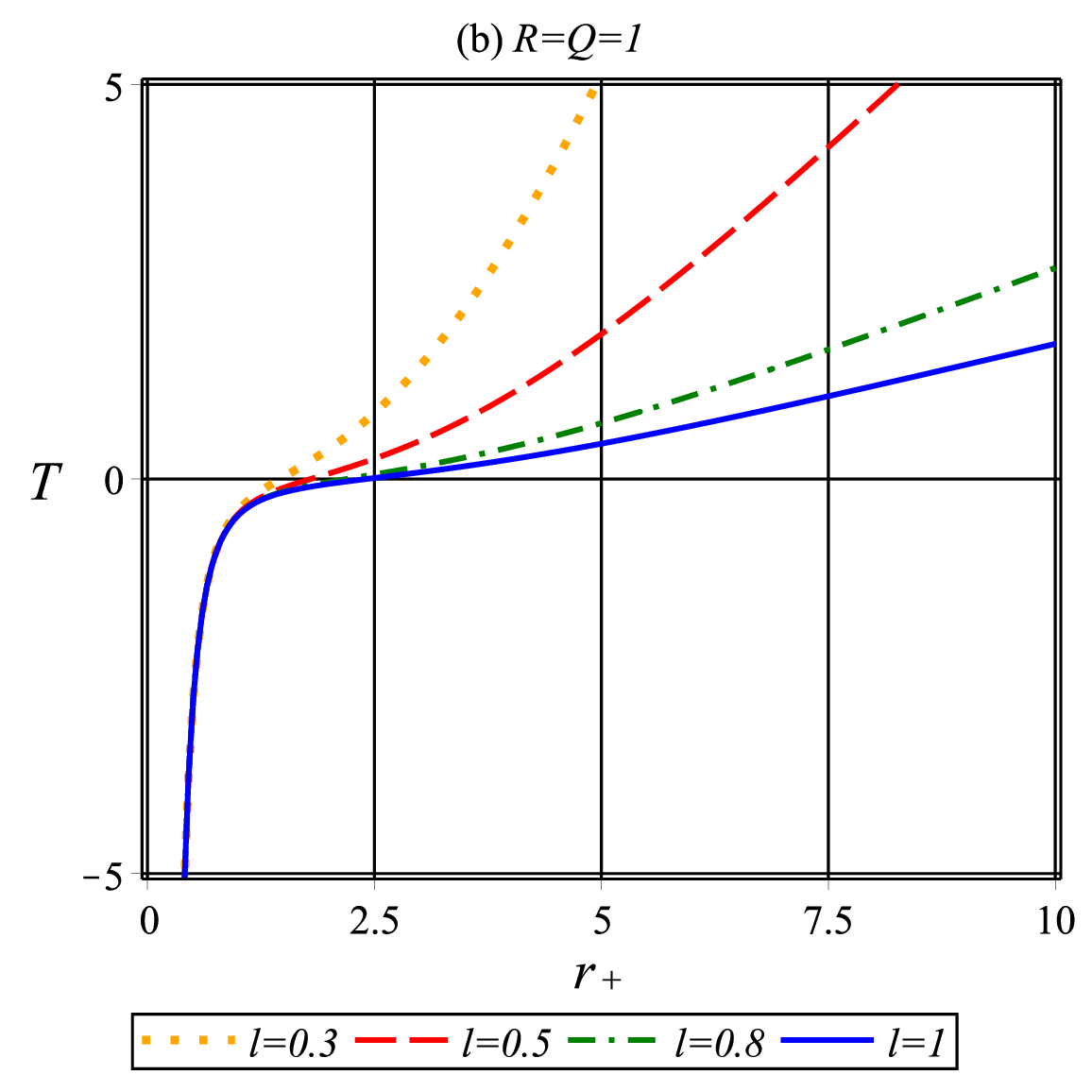}\\
\includegraphics[width=65 mm]{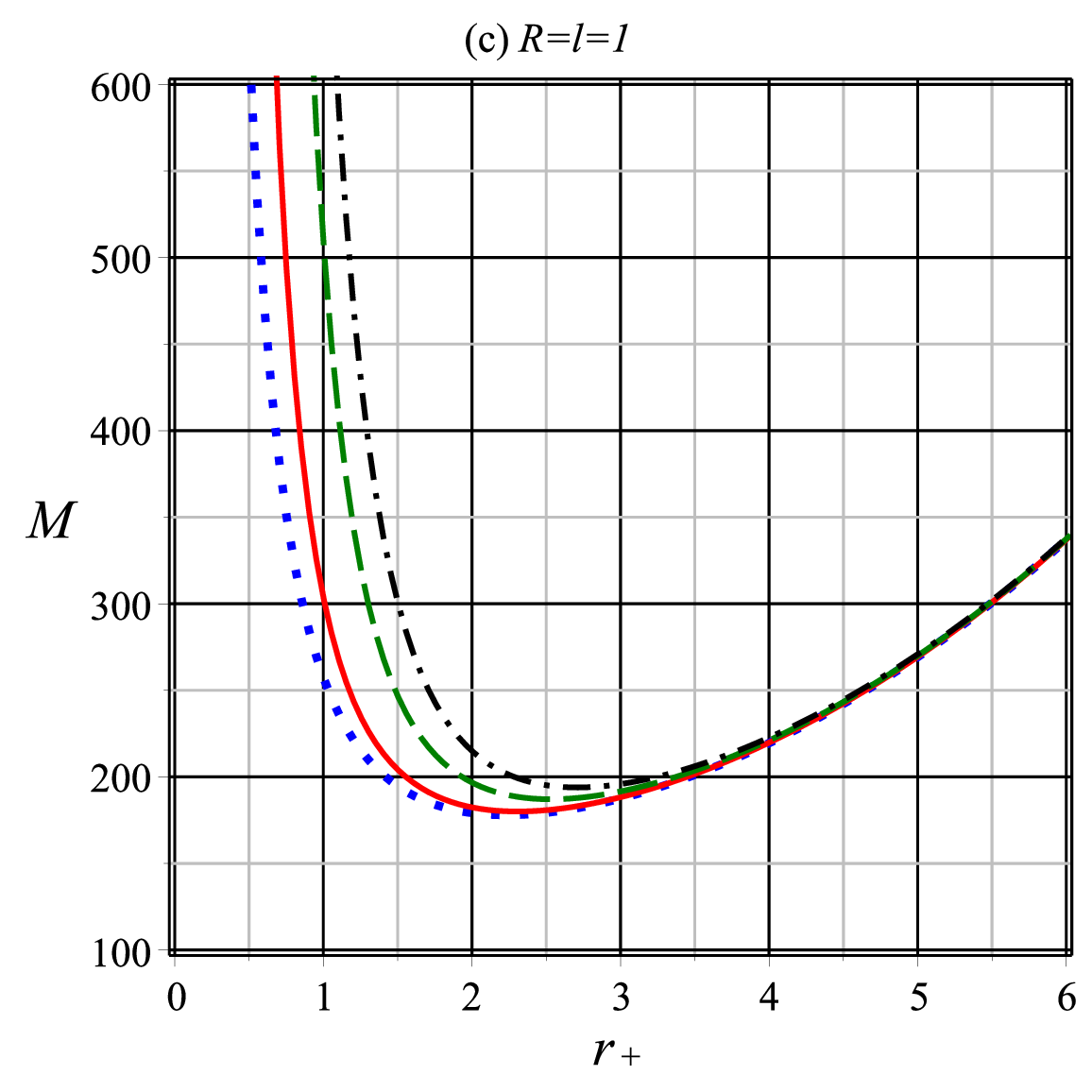}\includegraphics[width=65 mm]{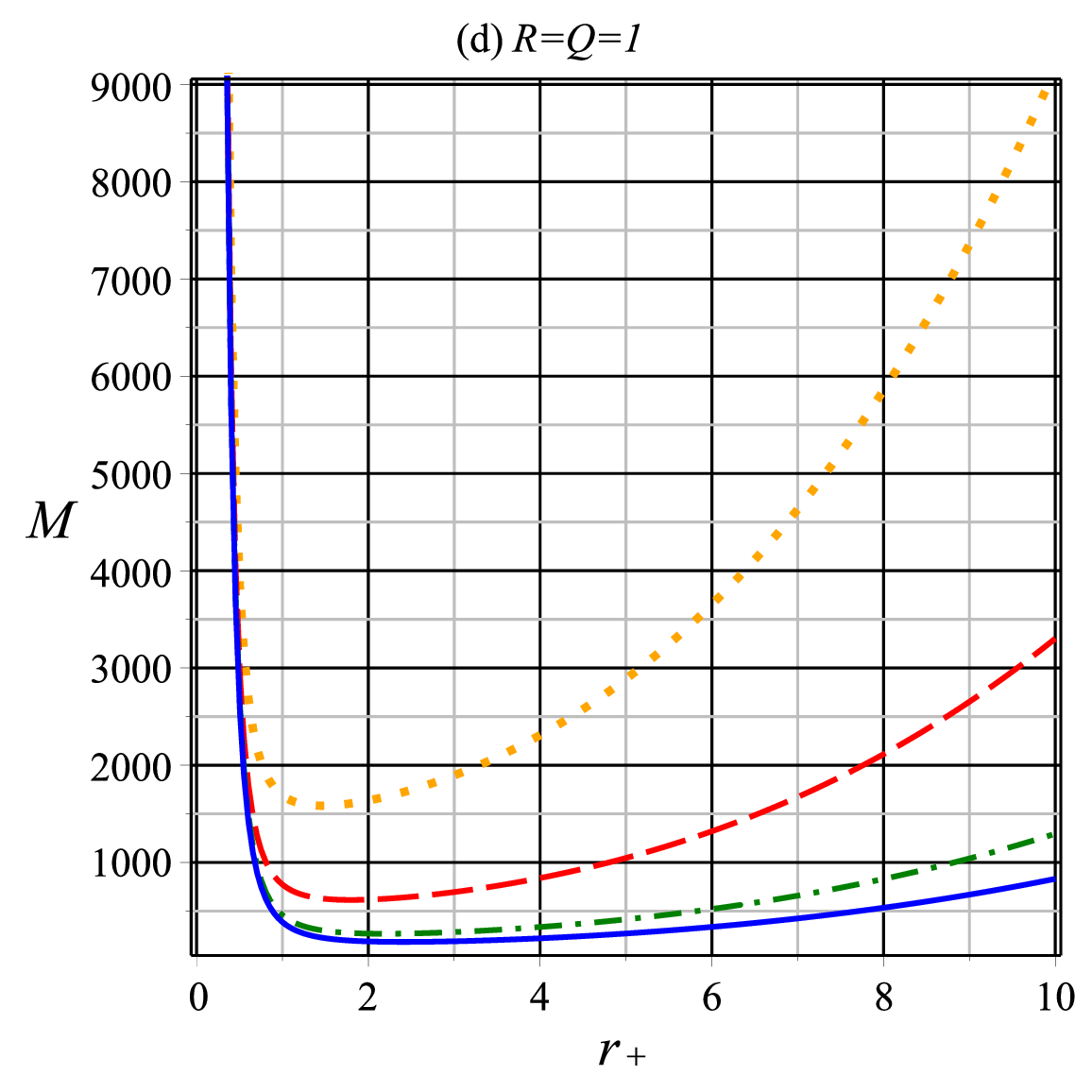}
 \end{array}$
 \end{center}
\caption{Temperature and Mass in terms of $r_{+}$.}
 \label{10}
\end{figure}

Using  the transformation like $V^{\frac{2}{3}}\rightarrow \mathcal{V}$, we can  write
\begin{equation}\label{Cosmol9}
\left(P-\frac{3}{16\pi L^{2}}\right)\mathcal{V}=1+\frac{\mathcal{B}}{\mathcal{V}}+\frac{\mathcal{C}}{a_{0}+b_{0}\mathcal{V}}
+\frac{\mathcal{D}\mathcal{V}}{(a_{0}+b_{0}\mathcal{V})^{2}}.
\end{equation}
It is like virial expansion, and  hence $\mathcal{B}$, $\mathcal{C}$ and  $\mathcal{D}$ are virial coefficients. Thus, even this corrected system behaves like a   Van der Waal fluid \cite{PV1, PV2, PV3, PV4}.  So, even though  the value of the specific terms is modified by the T-duality, the form of the equations, and hence the Van der Waals behavior still holds in the system corrected by T-duality.
\\
As the black hole size reduces  due to Hawking radiation, the  pressure also    reduces. Then the radius a  black hole reaches a critical value, and a black  remnant is  formed at
\begin{equation}\label{Cosmol10}
r_{min +}=\sqrt{\frac{\sqrt{96\pi^{2}R^{2}+L^{2}}-L^{2}}{6}}.
\end{equation}
It may be noted that at  this radius, the temperature, entropy and specific heat  of the black remnant vanish
\begin{eqnarray}
 T (r_{min +}) = 0, & S(r_{min +}) = 0, & C(r_{min +}) =0.
\end{eqnarray}
The formation of the black remnant can be seen from  Fig. \ref{88}. Now from  in  Fig. \ref{88} that a black remnant forms at    $M_{min}\approx1.25$. This is the mass of the black remnant, when a cosmological constant is added to the system.
\\
 It is also interesting to consider effects of chemical potential on this Van der Waal fluid like system. It   corresponds to adding an electric charge to the black hole \cite{c1, c2}. Now it may be noted that as electric charge is just a conserved charge due to internal symmetry of the system, so it cannot be modified by space-time symmetries. Thus, we do not expect any modification to the terms proportional to the electric charge in the  black hole. However, non-trivial effects can be obtained in this system by the modification  to the matter part \cite{a}, which occurs due to the T-duality of strings theory compactified on extra dimensions. Now we expect that this solution should reduce to  an AdS black hole modified by T-duality, if the electric charge vanishes. Thus, we can write the metric for such a black hole as
 \begin{equation}\label{Q1}
f(r)=1-\frac{2Mr^2}{(r^2+4\pi^{2}R^{2})^{3/2}}+\frac{r^{2}}{L^{2}}+\frac{Q^{2}}{r^{2}},
\end{equation}
where $Q$ is electric charge of this black hole corrected by T-duality. Now, it is possible to  write  the modified   temperature for this black hole as
\begin{equation}\label{Q2}
T=\frac{3r_{+}^{6}+L^{2}r_{+}^{2}(r_{+}^{2}-8\pi^{2}R^{2})-L^{2}Q^{2}(r_{+}^{2}+16\pi^{2}R^{2})}{4\pi L^{2}r_{+}^{3}(r_{+}^{2}+4\pi^{2}R^{2})}.
\end{equation}
It may be noted that the  Komar mass  is now given by
\begin{equation}\label{Q3}
M=\frac{(1+\frac{r_{+}^{2}}{L^{2}}+\frac{Q^{2}}{r_{+}^{2}})(r_{+}^2+4\pi^{2}R^{2})^{3/2}}{2r_{+}^{2}}.
\end{equation}
It is possible to write the first law of thermodynamics for a black hole with a charge and cosmological constant
\cite{c1,c2}. Even though the solution, and the value of the specific quantities would get modified at short distances, due to T-duality,  the first law of thermodynamics can still be written  as,
\begin{equation}\label{Q4}
dM=TdS+PdV+\Phi dQ,
\end{equation}
where $\Phi$ can be interpreted as chemical potential, and obtained using
\begin{equation}\label{Q5}
\left(\frac{dM}{dQ}\right)_{S, V}=\Phi.
\end{equation}
It is possible to calculate the Helmholtz free energy of the system, and observe that it get corrected by these non-perturbative corrections.
So,  we can write  this corrected   Helmholtz free energy for this system as
\begin{eqnarray}\label{Q6}
F&=&{\frac {9\pi^{2} R^{2}r_{+}}{4{L}^{2}}}-{\frac {3{\pi}^{2}R^{2}}{2r_{+}}}-{\frac {3{\pi }^{2}{R}^{2}r_{+}}{{L}^{2}}}-{\frac {{r_{+}}^{3}}{4{L}^{2
}}}+\frac{r_{+}}{4}-{\frac {12{\pi }^{4}{R}^{
4}r_{+}}{ \left( 4{\pi }^{2}{R}^{2}+{r_{+}}^{2} \right) {l}^{2}}}\nonumber\\
&-&{\frac {9{\pi }^{4}{R}^{4}r_{+}}{ \left( 4{\pi }^{2}{R}^{2}+{r_{+}}^{2}
 \right) {L}^{2}}}+{\frac {3{\pi }^{2}{R}^{2}r_{+}}{4{\pi }^{2}{R}^{2}
+{r_{+}}^{2}}}+{\frac {9\pi^{2} R^{2}r_{+}}{16{\pi }^{2}{R}^{2}+16{r_{+}}^{2}}}\nonumber\\
&+&{\frac {12{\pi }^{3}{R}^{3}}{{L}^{2}}\tan^{-1} \left({\frac {r_{+}}{2
\pi R}} \right) }-3\pi R\tan^{-1} \left({\frac {r_{+}}{2\pi R}}
 \right)\nonumber\\
 &+&\left(\frac{57}{16r_{+}}-\frac{3\pi^{2}R^{2}}{r_{+}^{3}}-\frac{21r_{+}}{16(4\pi^{2}R^{2}+r_{+}^{2})}+\frac{3\tan^{-1} \left({\frac {r_{+}}{2\pi R}}
 \right)}{\pi R}\right)Q^{2}.
\end{eqnarray}
It may be noted that as in the limit in which we the cosmological constant and electric charge becomes zero, the corrected Helmholtz free energy reduced to the the Helmholtz free energy of a Schwarzschild black hole.
Then, specific heat, at the first order approximation, in terms of event horizon radius, can be  obtained as,
\begin{equation}\label{Q7}
C=\frac{2\pi r_{+}^{2}(4\pi^{2}R^{2}+r_{+}^{2})(3r_{+}^{6}+L^{2}r_{+}^{2}(r_{+}^{2}-8\pi^{2}R^{2})-L^{2}Q^{2}(r_{+}^{2}+16\pi^{2}R^{2}))}
{3r_{+}^{8}+(36\pi^{2}R^{2}-L^{2})r_{+}^{6}+L^{2}(28\pi^{2}R^{2}+3Q^{2})r_{+}^{4}+(32\pi^{4}R^{4}+84\pi^{2}R^{2}Q^{2})L^{2}r_{+}^{2}+192L^{2}Q^{2}\pi^{4}R^{4}}.
\end{equation}
We observe  that    the  charged system is more stable than uncharged system. So, the chemical potential   increasing stability of such a system, even for a black hole corrected by T-duality. \\
Now we can write the  black remnant mass in presence of cosmological constant and chemical potential. We can obtain radius, where temperature of the black hole becomes  zero
$T(r_{min +}) =0$. It is illustrated by plots of the Fig. \ref{10} (a) and (b). However, it is possible to obtain analytical expression for   charged black hole   as
\begin{equation}\label{Q8}
r_{min +}^{2}\approx \frac{L}{6}\left[-L+\sqrt{L^{2}+12Q^{2}+96\pi^{2}R^{2}}\right],
\end{equation}
which is reduced to Eq. (\ref{Cosmol10}) for $Q=0$. The specific heat for a charged black hole is plotted in  the Fig. \ref{9}.
It may be noted from  Fig. \ref{10} (c) and (d), this system has a  minimum mass.  We can also observe      effects of   chemical potential and cosmological constant on it. From the Fig. \ref{10} (c), we observe  that the electric charge (hence chemical potential) increases this mass.
\\
So, apart from black remnants in the Schwarzschild  black solution \cite{a}, black remnants seems to also form in various other  black holes solutions,
modified by T-duality. Thus,  the formation of black remnants seems to be a general feature of such corrections.
In fact, as T-duality can lead to   ultraviolet finiteness of string theory \cite{s16, s18},   it is expected that it would modify various    black holes solutions   \cite{a, a1, a2}, and thus produce black remnants.
The existence of such black remnants could solve the information paradox, and the information can be stored in such a black  remnant, and this black    remnant would not evaporate \cite{bhin, bhin1}. It has been demonstrated that due to quantum effects information need not be additive, and so  a small number of quanta can contain a large amount of information \cite{info}. Thus, the last stages of the black holes, when they form a black remnants can contain the entire information of a black hole. There is a cosmological limit on the size and total information of a  large  black hole. So, taking the Lloyd's estimate,   the total information of the universe is around   $10^{120}$ bits \cite{info1}, and this is the maximum information that can be contained in a black hole. It has been demonstrated that all this information can be stored in a black remnant which is only $40$ Planck masses \cite{info}.
Thus, the formation of such a black remnant  has to potential to resolve the information loss paradox. Here we have demonstrated that such a resolution to the information loss paradox can occur due to non-pertubative corrections to black hole solutions.
It is also interesting to note that such black remnants can also form primordial black hole, as these primordial black holes would evaporate to such black remnants. Such primordial black hole remnants can form are a candidate for cold dark matter as they can form weakly interacting massive particles \cite{pn12, pn14}. Thus, the modification of black holes from T-duality can also be used to motivate the production of primordial black hole remnants, which can in turn be used as a candidate for cold dark matter.
This is because it has been suggested that primordial black hole remnants can act cold dark matter \cite{prim}. Thus, the primordial black holes could have evaporated to form primordial black hole remnants, due to non-pertubative string theoretical effects, and these in turn could account for cold dark matter.

\section{Conclusion}
In this work, we have analyzed non-perturbative   string theoretical corrections to black holes, which occur from the finiteness of  string theory. Such finiteness of string theory occurs due to  T-duality \cite{green1, green01, green02, green2}, as according to T-duality, the behavior of string theory below string length scale is the same as its behavior above it. This finiteness modifies the short distance behavior of black holes, which in turn changes the physics of such black holes.
It has been observed that the T-duality modifies a  four dimensional  Schwarzschild black hole to a  Bardeen black hole \cite{a},
and the Bardeen black holes have a black remnant state with zero temperature \cite{bardeen, bardeen1}.  In this paper, we have first  studied the thermodynamic stability of such black remnants using Gibbs free energy. Furthermore, we have generalized these results to higher dimensional black holes. It was argued that the formation of such  black remnants due to T-duality could  be used to explain the absence of  mini black holes at the LHC.  Furthermore, we were able to obtain bounds for the zero point length (related to the invariance of the theory  under T-duality), from the LHC data. We also obtained such a bound for future accelerators with energies around $100 \, Tev$  \cite{LHC2, LHC4}.
\\
We also analyzed the modification to other black hole solutions from T-duality. Thus, we analyzed such modification to an AdS black hole. We identified  the cosmological constant in AdS black hole solution, with the pressure of the system \cite{PV1, PV2}, and  analyzed the effect of such a modifications on such a black hole solution. It was observed that this corrected black hole solution still had a Van der Waals behavior. Finally, we have analyzed such corrections occur due to T-dualtiy, to a black hole with a chemical potential. It was observed that such modification produced zero temperature black remnants. As  such modifications changed the last stages of the evaporation of black holes, it was argued that  they  can have important consequences for the black hole information paradox. This is because a small number of quanta can contain a large amount of information \cite{info}, and so black remnants can contain all the information of a black hole. It was also argued that these  black remnants can also form for primordial black holes, which  could be  a candidate for cold dark matter \cite{prim}. We have analyzed several interesting   physical consequences of these black remnants, which form due to non-perturbative string theoretical effects.
\\
It may be noted that it was expected that such black remnants  would form due to string theoretical effects, due to the ultraviolet finiteness of string theory \cite{s16, s18}. In fact,
  a  four  dimensional magnetic black hole solutions of heterotic string theory  had  been  constructed using $SU(2)/Z(2Q+2)$ WZW orbifold, and it had also been demonstrated  certain  marginal operators can deform this theory to an asymptotically flat black hole. It was   demonstrated  that   black remnants can form in this system \cite{r1}. Thus, black remnants  would be expected  to form from  string theoretical effects. However, in this paper, we have  constructed and analyzed such solutions, using the modification to matter in black holes \cite{a},  which occurs due to the T-duality in string theory \cite{green1, green01, green02,  green2}.
\\
It would be interesting to analyze such modification to the thermodynamics of various geometries that occur in string theory.
This can be done by analyzing various solution to different supergravity actions, and then using analyzing this non-perturbative  corrections to such supergravity solutions. It may be noted that such modifications are produced by non-perturbative effects,
and cannot be produced by pertubative calculations. This is because the modified Green's function does not diverge, but if it is expressed as a pertubative sum, each term would tend to diverge \cite{green1, green01, green02,  green2}.  So, to analyze such modifications of supergravity solution, it would be important to first obtain a perturbative   black hole solution, and then analyze such  non-pertubative corrections to that solution.
One of the important differences between this non-pertubative modification to a black hole solution, and the usual black hole solution is that this black hole solution is free from singularities. Now singularities are also important in cosmology, and it might be possible to remove the big bang singularity, by modification of different cosmological solutions, using this formalism.
It would be interesting to analyze the constraints on such cosmological models from different cosmological data sets.\\\\
\section*{Acknowledgement}
We would like to thank Mubasher Jamil
for his contributions and discussions in the development of the earlier part of this work.

\end{document}